\documentclass[aps,prd,preprint,groupedaddress,nofootinbib,floatfix]{revtex4-1}
\renewcommand{\thetable}{\arabic{table}} 
\usepackage{amsmath}
\usepackage{graphicx}
\usepackage{amsbsy}
\usepackage{amssymb}
\usepackage{palatino,avant,graphicx,color}
\usepackage[none]{hyphenat}
\usepackage{xy}
\usepackage{changes}
\usepackage{wrapfig}
\usepackage{float}
\usepackage{tabularx}
\usepackage{tikz}
\def\checkmark{\tikz\fill[scale=0.4](0,.35) -- (.25,0) -- (1,.7) -- (.25,.15) -- cycle;} 

\usepackage{hyperref}
\usepackage{mathtools}
\makeatletter
\newcommand{\printfnsymbol}[1]{%
  \textsuperscript{\@fnsymbol{#1}}%
}
\makeatother

\begin{document}

\title{Thin liquid film as an optical nonlinear-nonlocal medium and memory element in integrated optofluidic reservoir computer}
  \author{Chengkuan Gao}
  \author{Prabhav Gaur}
  \author{Shimon Rubin}
 \email{rubin.shim@gmail.com}
 \author{Yeshaiahu Fainman}
 \affiliation{Department of Electrical and Computer Engineering, University of California, San Diego, 9500 Gilman Dr., La Jolla, California 92023, USA}


\begin{abstract}

Understanding light–matter interaction enables harnessing physical effects to translate into new capabilities realized in
modern integrated photonics platforms.
Here, we present the design and characterization of optofluidic components in integrated photonics platform, and computationally predict a series of novel physical effects which rely on 
thermocapillary-driven interaction between waveguide modes to topography changes of optically thin liquid dielectric film.
Our results indicate that this coupling introduces substantial self-induced phase change and transmittance change in a single channel waveguide, transmittance through Bragg grating waveguide and nonlocal interaction between adjacent waveguides.
We then employ the self-induced effects
together with the inherent built-in finite relaxation time of the liquid film, to demonstrate that the light-driven deformation can serve as a reservoir computer capable to perform digital and analog tasks, where the gas-liquid interface operates both as a nonlinear actuator and as an optical memory element.

\end{abstract}

\maketitle

\section*{Introduction}
\sloppy

Light–matter interaction is central to advancing our
understanding of the properties of both light and matter.
Emergence of robust integrated photonic platforms over the last two decades, characterized by miniaturized cross-section and higher refractive index contrast, 
were leveraged to enhance the optical intensity thus achieving nonlinear response of various matter degrees of freedom. 
More recently, nonlinear integrated photonics platforms \cite{Borghi2017} demonstrated promising capabilities to conduct 
basic scientific research in technologically attractive applications such as frequency conversion and third-harmonic generation \cite{Corcoran2009}, supercontinuum generation \cite{Gaeta2019}, emerging quantum photonics applications \cite{Silverstone2014}, and is considered as an attractive platform for future non-conventional computation architectures \cite{VanDerSande}.
In particular, efficiency of neuro systems to process computational tasks which are challenging to traditional Turing - von Neumann machines \cite{Hasler2013} 
due to sequential 'line-by-line' operation,
inspired development of machine learning based Neuromorphic Computing (NC) and Recurrent Neural Network (RNN) computational models \cite{Jaeger2001,Maas2002} as well as its subset called Reservoir Computing (RC) \cite{Verstraeten2007,Luko2009}.
Similarly to RNN, the recurrent signal in RC constitutes a memory capable to participate in the parallel computation; 
however, in contrast to RNN, where computationally complex algorithms are required to tune internal weights in the network, in RC the dynamics occurs in a fixed recurrent network and only external weights are digitally tuned thus leading 
to a significant reduction of the training computation time and the size of the required memory.
Consequently, various physical systems can in principle serve as powerful RC platforms  \cite{Tanaka2019} where the complex and nonlinear signal output produced by the physical system is collected and used
for supervised learning in order to obtain the desired set of weight coefficients during the digital learning stage.
Key properties physical RC systems should satisfy include sufficiently large reservoir dimensionality allowing to separate features of distinct states according to their dynamics,
short term (fading) memory, and a balance between sensitivity to separability.
Among the different proposed physical mechanisms, optical-based systems \cite{Wetzstein2020, Pruncal2017} are particularly attractive due to their inherent parallelism, speed of light data propagation and processing, and relatively low energy consumption leading to both on-chip \cite{Vandoorne2014,Tait2017,Peng2018} and free space \cite{Larger2017,Rafayelyan2020} realizations.

In this work we employ the recently theoretically proposed light-liquid nonlinear and nonlocal interaction mechanism \cite{Rubin2018,Rubin2019}, where localized light-induced heating invokes thermocapillary (TC) \cite{Marangoni1871, Pearson1958, Levich1962} driven deformation of optically thin gas-liquid interface \cite{Wedershoven2014,RubinHong2019}, to predict and characterize a set of new nonlinear-nonlocal effects in integrated components, and then leverage these effects to achieve RC capabilities.
\begin{figure}[t]
	\includegraphics[scale=0.21]{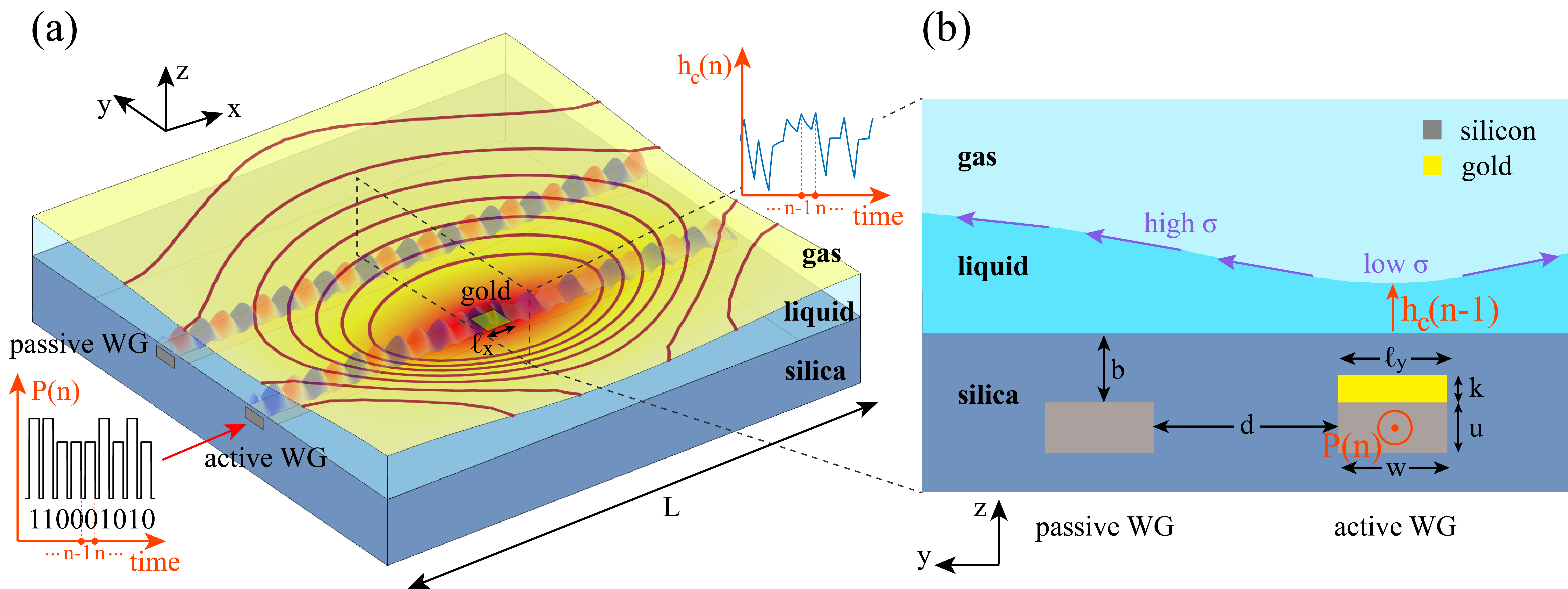}
   \caption{Schematic description of the key components in the integrated optofluidic system under study describing the underlying mechanism of the light-liquid interaction and the memory used for RC. 
   (a) 3D perspective presenting box-shaped liquid cell of length $L=10$ \text{$\mu$}m bracketed by vertical silicone walls, and an active Si channel WG on silica substrate covered with a gold patch of dimensions $\ell_{x} \times \ell_{y} \times k$ enabling light-induced heating, subsequent heat transport to the gas-liquid interface which in turn triggers surface tension gradients, represented by blue arrows in (b), leading to TC effect and to liquid film thinning. 
   The latter leads to a self-induced phase change \textcolor{black}{and/or to transmittance change} in the single WG setup, and also to a nonlocal effect where the phase in the adjacent passive WG is modified due to liquid deformation.
   Darker colors on the liquid's surface in (a) correspond to higher temperature whereas the dark red lines denote equal height levels.
 A sequence of high and low optical power pulses correspond to logic '0' and '1', respectively. 
 The finite relaxation time of the gas-liquid interface 
 allows to employ it as a short term memory where $n-1$-th pulse of power $P(n-1)$ induces $h_{c}(n-1)$ liquid thickness affecting the subsequent $n$-th pulse of power $P(n)$.
   (b) 2D normal section, consisting of Si channel WGs of width $w = 500$ nm and height $u = 220$ nm buried in depth $b = 80$ nm under silica top surface and hosting a gold patch of thickness $k = 20$ nm and in-plane dimensions $\ell_{x} =\ell_{y} = 500$ nm; the initial liquid depth is $h_{0} = 500$ nm. 
   }
    \label{Scheme}
\end{figure} 
Fig.\ref{Scheme} presents schematic illustration of the typical setup
where the optical mode which propagates in the waveguide (WG) covered with thin liquid film, invokes heating in a small gold patch on the top facet of the WG due to optical absorption, leading in turn to TC effect and to liquid deformation. 
In case the liquid film is sufficiently thin, liquid deformation modifies the overlap of the evanescent tail with the gas phase above the liquid film, leading to self-induced change of optical mode properties such as phase, \textcolor{black}{intensity} or coupling coefficient to another WG, which we consider below.
For numerical simulations we define
silicon (Si) or silicon-nitride (SiN) 
channel WGs designed to carry single mode at wavelength $1550$ nm (we consider TM polarization), which is buried in silica (SiO$_{2}$) layer in order to provide a flat surface for more straightforward numerical implementation. 
Our extensive 3D multiphysics \cite{comsol} numerical analysis which takes into account the full dynamics of various degrees of freedom, allows quantitative analysis beyond previous studies which considered limiting assumptions such as small liquid deformation and effective 2D approximation schemes of surface plasmon polariton \cite{Rubin2018} and slab WG modes \cite{Rubin2019}.
In particular, our simulations include intricate coupling where optical dissipation in the gold patch 
leads to Joule heat generation and its transport to the gas-liquid interface; the latter triggers surface tension gradients along the gas-liquid interface which in turn invoke TC flow accompanied by liquid's thickness change above the active WG (Fig.\ref{Scheme}b) (\textcolor{black}{in our terminology active/passive WG corresponds to presence/absence of gold patch}).
For the simplest case of a single active WG the corresponding deformation of the dielectric cladding leads to a self-induced change of the accumulated optical phase thus constituting a two-way interaction where the optical mode affects itself through the liquid, provided the gas-liquid interface overlaps with the evanescent tail. 
Furthermore, substantial in-plane spatial scale of liquid's indentation allows to achieve the so-called optical nonlocality where the active optical mode affects the passive optical mode even if the passive WG is displaced from the region of maximal optical intensity. 
We then employ the self-induced nonlinear phase change in a single WG or alternatively the nonlocal coupling change between adjacent WGs, as a nonlinear actuator which provides internal feedback and also exhibits memory capability due to finite relaxation time of the gas-liquid interface, allowing to perform digital and analog RC tasks. 

\section*{Methods}

\subsection{Details of multiphysics simulation} 

To capture the complex coupling between light propagation, fluid dynamics, heat transport and surface tension effects, we employed finite-element based commercial software COMSOL Multiphysics$^\circledR$ \cite{comsol} with Wave Optics, heat transfer and CFD module where 
changes of thin liquid film geometry are simulated by the moving mesh method. 
Since electromagnetic dynamics occurs on much shorter time scales compared to other processes we consider, the optical simulation is performed in the frequency domain whereas all other processes are simulated in the time domain. 
For the WG and the thinner gold patch we use free tetrahedral mesh 
resulting in approximately $600$K mesh elements and $4$ million degrees of freedom. 
These memory demanding simulations are performed on a constructed dedicated server with $16$ cores and $512$ GB memory; for instance single self-induced phase change simulation presented in Fig.\ref{SelfInducedPhase} takes more than $24$ hours simulation time. 
The governing equations, matching and boundary conditions as well as key parameters used for the simulation are provided in supplemental material (SM). 
In all simulations we assume the initial liquid surface is flat and forms contact angle of value $90^{o}$ with the vertical silica boundary walls.

\subsection{Details of RC simulation}

Construction of the computational scheme for optofluidic RC involved three main steps. 
In the first step we employed approximately $100$ 3D simulations each providing gas-liquid interface deformation under equivalent heat source (see SM Fig.S4a) for a 
variety of different developed temperatures in the gold patch and initial configurations (see SM Fig.S3).
In the second step we extracted the corresponding data describing evolution of $h_{c}(t)$ and performed linear fitting to derive the following reduced phenomenological 1D model for $h_{c}(t)$ 
\begin{equation}
    \dot{h}_{c}(t) = \alpha(P) \cdot h_{c}(t) + \beta(P),
\label{NModelH}
\end{equation}
with power dependent coefficients $\alpha(P), \beta(P)$ (see SM Fig.S5b).
\textcolor{black}{While the first step enabled us to generate sufficiently large dataset in a reasonable time, the second step allowed to reduce computation time of the RC algorithm shortly described below.}
Finally, we implemented RC scheme \cite{Luko2012} by using Matlab (R2019b, Mathworks, Natick, MA) \cite{Matlab} to simulate the optical power at the detectors in MZI circuit, described in Fig.\ref{XOR}a, as a function of the exciting signal. 
In particular, we employed Euler method in order to integrate Eq.\ref{NModelH} with time step of $0.1$ ms and employed the relation Fig.S5c to derive the corresponding phase shift $\Delta \varphi_{TC}$ and the accompanying optical signal at the detectors.
To implement RC by employing our system we first obtain training matrix, without tuning internal system properties, and then use this matrix to perform test stage which performs relevant computation.
In particular, to accomplish the training stage we first collected the output data in $D_{1,2}$ and arranged it in $N p_{w} \times 2$ dimensional matrix where $N$ is the number of bits used for training and $p_{w}$ is the excitation time (which affects the number of rows in the output matrix).
For instance, in our XOR simulation for training step we used $p_{w} = 25$, $N=1000$.
We then rearrange the output matrix in matrix $X = (M, N)$, where $M = 2 p_{w}$ and each column in the matrix $X$ represents data from time step $i$ ($i=1,...,N$), 
and use it to solve the following linear equation for the weight matrix $W_{out}$
\begin{equation}
    Y = X \cdot W_{out},
\end{equation}
where $Y$ is a pre-determined desired output.
Here, $W_{out}$ is the $M \times 1$ dimensional vector that is determined by employing Tikhonov (ridge) regression, which is expected to be executed digitally in future experimental realizations, via Matlab's built-in 'ridge' function, which minimizes the following expression
\begin{equation}
    \vert Y - Y_{T} \vert^{2} + \beta \vert W_{out} \vert^{2},
\label{Norm}    
\end{equation}
with $\beta$ serves as a regularization parameter and $\vert ... \vert$ denotes the Euclidean norm.
Note that the first term in Eq.\ref{Norm} penalizes large differences between the output vector $Y$ and the desired output $Y_{T}$, whereas the second term penalizes large weight values which facilitates better performance \cite{Luko2012}. 
The corresponding closed form for $W_{out}$ is then given by
\begin{equation}
    W_{out} = \left( X^{T} X + \beta  I_{N} \right)^{-1} X^{T} Y,
\end{equation}
where non-zero $\beta$ multiplies a unity matrix $I_{N}$ of dimension $N$ thus adding nonzero values to the diagonal of matrix $X^{T} \cdot X$ and potentially regularizing it.
To implement test stage we feed previously unseen driving sequence, generate new design matrix $X_{test}$ and operate on it with the previously derived $W_{out}$ via
\begin{equation}
    Y_{test} = X_{test} W_{out}
\end{equation}
To finalize the computation result the output vector $Y_{test}$ is subjected to threshold step with values above (below) $0.5$ being rounded to one (zero).
The corresponding error rate ($Er$) is then determined by comparing the computation result after the threshold with the true result, leading to
\begin{equation}
    Er = \dfrac{100 \times \gamma}{N_{test}},
\end{equation}
where $N_{test}$ is the total number of bits in the test sequence and $\gamma$ is the number of errors in the computation.
For test stage of the XOR task we employed the values $N_{test}=20$ and $M=50$, whereas for 'zero'/'one' handwritten digits classification we employed 
$N=12,665$, $N_{test}=2,115$ and $M = 28 \times 28 \times 2 = 1,568$ weights.
The corresponding ridge parameter value we used to train all models in our work was $0.01$, yet determining its optimal value or region of applicability allowing even better performance would require employing ridge parameter estimation methods \cite{Chatterjee2013} which is more data science oriented investigation and thus beyond the scope of this work.

\section{Results}

\subsection{Nonlinear self-induced phase change}

\begin{figure}[t]
	\includegraphics[scale=0.27]{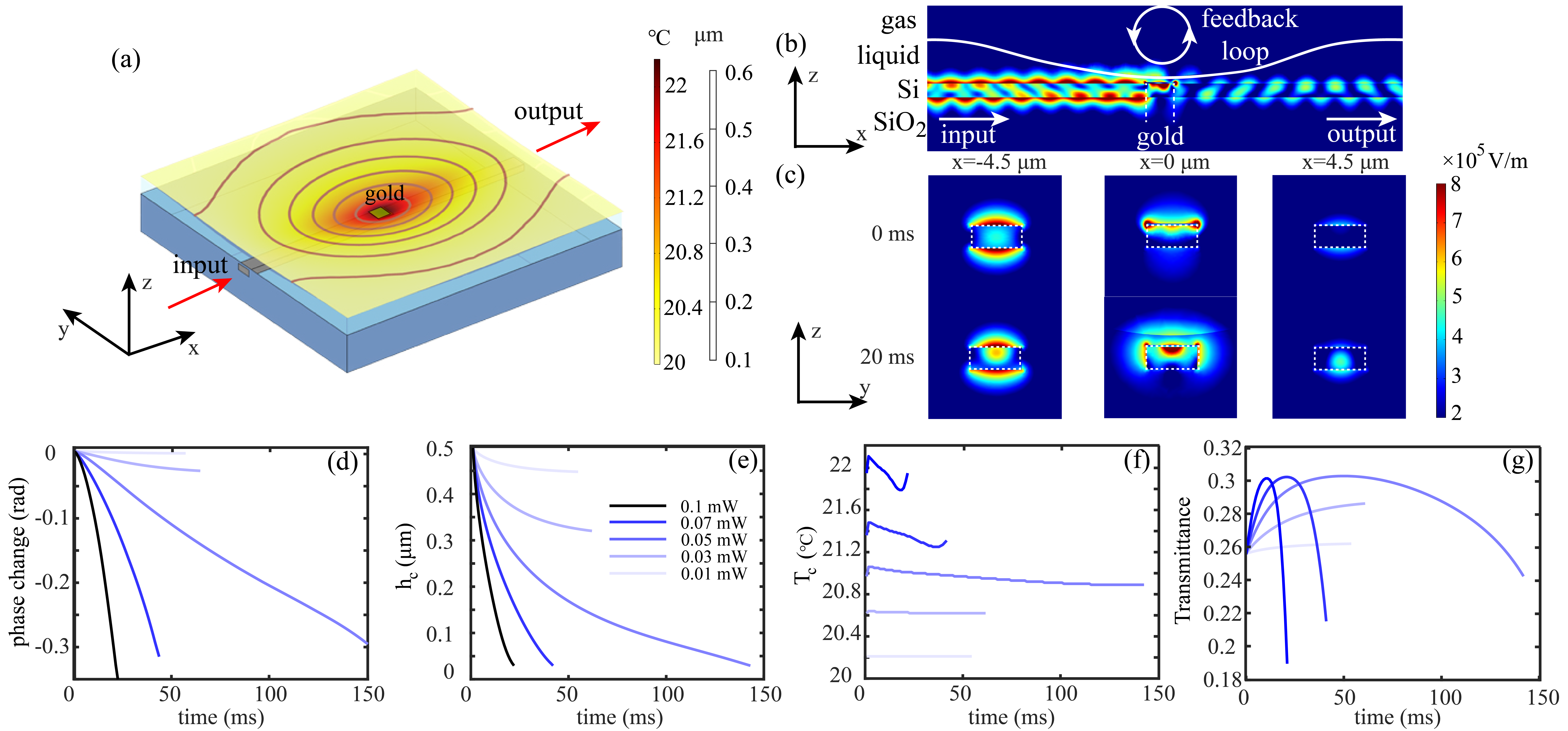}
   \caption{Numerical multiphysics simulation results presenting self-induced phase change in a single active WG covered with a thin liquid film. (a) Deformed gas-liquid interface under $0.1$ mW \textcolor{black}{CW light-induced TC effect at time $t=20$ ms (where $t=0$ s is the time moment when the optical mode began to propagate)}
   and (b) graphic representation of the internal optical-generated feedback (circular arrow) due to deformation of the gas-liquid interface, allowing to invoke transient dynamics serving as memory storing the previous optical pulse and use the dynamics to affect the deformed profile and the optical pulse at next time moments at the same actuation region. 
   \textcolor{black}{The colormap describes electric field magnitude along the $x-z$ plane at time moment $t=20$ ms; the gold patch resides on the top facet of the WG and its dimensions along the $x$-axis are bracketed by the white dashed lines.}
   \textcolor{black}{(c) Electrical field magnitude along the normal section $y-z$ at $x=-4.5$ $\mu$m, $x=0$ $\mu$m, and $x=4.5$ $\mu$m corresponding to left, central and right columns, respectively, under $0.1$ mW incident intensity; first and second rows correspond to mode profiles at initial time $t=0$ ms and $t=20$ ms, respectively.}
   (d) The corresponding phase change as a function of different optical powers in the WG due to 
   (e) liquid film thickness change, where $h_{c}$ is liquid thickness above the center of the gold patch.
   \textcolor{black}{(f) The underlying change of the average temperature of the gold patch, $T_{c}$, and (g) is the corresponding transmittance as a function of time.}
   Key parameters: initial temperature $20^{o}$; wavelength of TM mode is $1550$ nm; refractive index of liquid is $1.44$ (same as silica substrate); geometric WG parameters are presented in Fig.\ref{Scheme}.}
    \label{SelfInducedPhase}
\end{figure}

Consider a single Si channel WG with integrated gold patch, covered with thin liquid film described in Fig.\ref{SelfInducedPhase}a.
Following the mechanism described above, the gold patch serves as an optical heater which induces TC-driven flows and deformation of the liquid dielectric film.
In our simulations we set liquid properties close to silicone oil which is an optically transparent and highly non-volatile liquid thus appropriate for future experimental realizations; see Supplemental Material (SM) Table 1 for simulation parameters.

Fig.\ref{SelfInducedPhase}a presents simulation results of the self-induced phase change effect due to TM mode propagating in a single active WG covered with a thin liquid dielectric film of initial thickness $500$ nm. 
In particular, Fig.\ref{SelfInducedPhase}a presents the corresponding deformation and temperature field of the gas-liquid interface at time $t=20$ ms \textcolor{black}{after the $0.1$ mW continuous wave (CW) mode began to propagate in the WG}, whereas
Fig.\ref{SelfInducedPhase}b presents 
the gas-liquid deformation as a feedback response (circular arrow) allowing to store information and affect subsequent optical pulses at the same compact spatial region, without the need of dedicated large footprint optical feedback elements, e.g., optical rings \cite{Mesaritakis2013}.
\textcolor{black}{The colormap describes electric field intensity at $t=20$ ms and thin liquid-deformation along the $x-z$ plane where $x$ is the propagation direction; note the effect of light reflection from the gold patch leading to distortion of the incoming mode.}
\textcolor{black}{Fig.\ref{SelfInducedPhase}c presents the electric field intensity at two different times along the planes $x=-4.5$ $\mu$m, $x=0$ $\mu$m and $x=4.5$ $\mu$m, where $x=0$ $\mu$m is the center of the gold patch. 
Specifically, the images at the first row indicate that at $t=0$ ms, i.e., prior to liquid film deformation, the incident TM mode experiences reduction in intensity but preserves its TM polarization. 
However, once the liquid film begins to deform, the second row indicates that after $20$ ms the incident mode changes its shape due to superposition with the reflected wave from the gold patch, and the transmitted mode is no longer top-bottom symmetric.
Fig.\ref{SelfInducedPhase}d,e,f,g present, the self-induced phase change, deformation of the thin liquid film above the gold patch center $h_{c}(t)$ (i.e., at region of minimal thickness just above the center of the gold patch), average temperature of the gold patch $T_{c}(t)$, and the corresponding transmittance as a function of time, respectively, under various optical powers.}
Naturally, increasingly higher power levels 
and accompanying temperature gradients, lead to more significant $h_{c}(t)$ and the corresponding self-induced phase change effects.
\textcolor{black}{Interestingly, at optical powers $0.1$ mW, $0,07$ and $0.05$ mW the corresponding liquid film deformation induces also prominent transmittance change leading to a power reduction by approximately a factor of $2/3$, which we attribute to impedance mismatch created by the liquid deformation.}
Notably, owing to the relatively low power levels required to activate TC-driven film thinning as well as the large refractive index contrast across the gas-liquid interface, sufficiently thin liquid film with high overlap of the evanescent optical tail with the gas phase supports few orders of magnitude higher self-induced phase change compared to that of \textcolor{black}{thermo-optical (TO)} effect.
In particular, Fig.S1 in SM indicates that the system described in Fig.\ref{SelfInducedPhase}a admits approximately three orders of magnitude higher self-induced phase change compared to a similar system where instead of liquid film one employs a dielectric solid material which doesn't support TC effect.
  It is worth mentioning that due to the moving mesh method employed in the multiphysics software \cite{comsol}, the simulation terminates once the gas-liquid interface reaches few elements thick film as measured from the bottom of the liquid cell, leading to phase change $\sim 0.3$ rad. 
  In future experimental realizations, which are not subject to such limitation, we expect to obtain even higher values of self-induced phase change and nonlocality scales, as well as self-induced transmittance/reflection modulation, shortly described below. 
 Furthermore, given the fact that liquid deformation is mainly concentrated around the compact gold patch, based on our study the size of the liquid cell can be in principle reduced to $2$ \text{$\mu$}m, whereas higher values of self-induced phase change can be achieved by placing several gold (or other metal) patches along the WG.

\subsection{Nonlocal effect}

\begin{figure}[t]
	\includegraphics[scale=0.2]{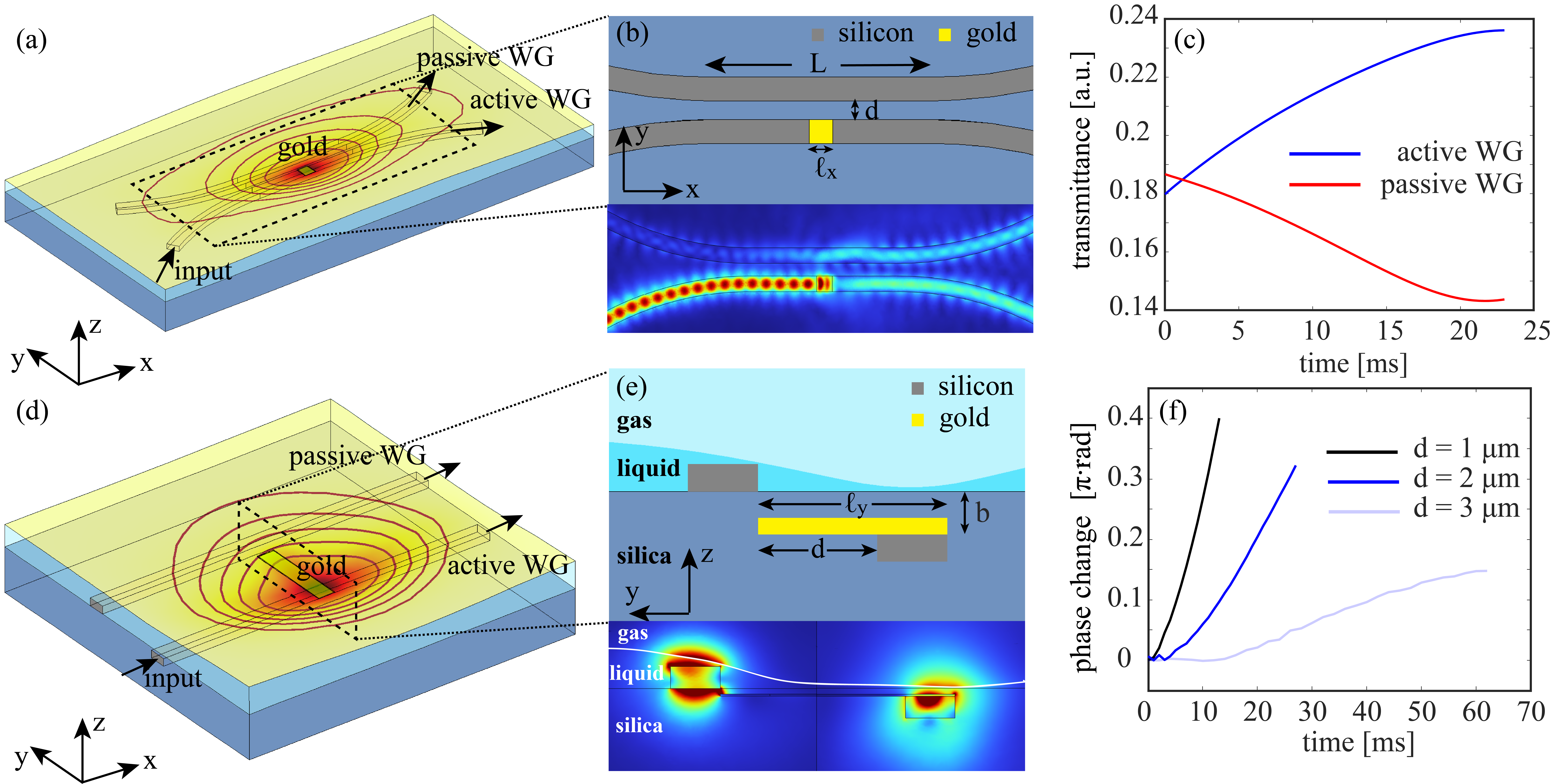}
   \caption{Numerical results demonstrating self-induced nonlocal interaction between two adjacent WGs in the close (a-c) and far (d-f) separation regimes, due to continuous TM mode of wavelength $1550$ nm injected to the active WG leading to TC-driven deformation of the liquid film; geometric parameters of WGs and the gold patch are provided in Fig.\ref{Scheme}. 
   (a-c) Close separation regime with directional coupler geometry of $d =0.4$ $\mu$m and interaction length $L=4.7$ $\mu$m. 
   (a) Temperature field and the geometry of the gas-liquid interface at time $t=23$ ms under $0.1$ mW; 
   (b) top view of the WGs with the gold patch of longitudinal dimension $\ell_{x} = 0.5$ \text{$\mu$}m \textcolor{black}{and the corresponding optical field at $t=23$ ms};
   (c) numerical simulation results of optical intensity change in both WGs as a function of time.
   (d-e) Far separation regime presenting three cases of parallel WGs separated by distances $d=1 ,2, 3$ $\mu$m and corresponding gold patches of lateral dimensions $\ell_{y} = d + w$. 
   (d) Temperature field and the topography of the gas-liquid interface at time $t=27$ ms under $0.1$ mW for the case $d=2$ \text{$\mu$}m;
   (e) side view of the WGs with elongated gold patch along the transverse direction, $\ell_{y}$, to facilitate heat transport from active WG to passive WG \textcolor{black}{and numerical results of two simulations at time $27$ ms where the mode is injected into active or passive WGs, indicating that the gold stripe doesn't support propagation of surface plasmon polariton between the WGs}; 
   (f) phase change in the passive WG mode as a function of time for three different distances from the active WG.}
    \label{NonLocal}
\end{figure}

Consider the optofluidic components schematically presented in Fig.\ref{NonLocal}, and analyze the effect of the self-induced TC-driven deformation on modes in the two WGs as a function of their distance $d$. 
In particular, we consider close and far separation regimes characterized by $d<L_{e}$ and $d>L_{e}$, respectively, where $L_{e}$ is the corresponding evanescent scale of the TM mode. 
To investigate the close separation regime we focus on the  directional coupler geometry \cite{Lifante2003} with minimal distance $d = 400$ nm presented in Fig.\ref{NonLocal}(a,b), whereas in the far separation regime we consider parallel WGs geometry presented in Fig.\ref{NonLocal}(d,e).
Fig.\ref{NonLocal}c presents simulation results of TM mode \textcolor{black}{injected only into the active WG}, indicating decrease of the coupling coefficient as a function of liquid deformation, leading to decreased (increased) transmittance of the TM mode through the passive (active) WG as a function of time.
In case the distance between the WGs is several microns ($d =$ $1 ,2 ,3$ \text{$\mu$}m), we can increase the nonlocal effect of the active WG on the passive WG by increasing the transverse dimension of the gold patch $\ell_{y}$ which facilitates more efficient heat transport and enables liquid film deformation profile which extends towards the passive WG.
This approach also allows to alleviate simulation limitation taking place due to its termination once the liquid film reaches thickness of few elements. 
\textcolor{black}{Fig.\ref{NonLocal}d presents 3D thin liquid film deformation and the underlying temperature map for $d=2$ $\mu$m case whereas Fig.\ref{NonLocal}e presents normal cross section with computational result of the corresponding modal structure.
In particular, the latter presents numerical results at time $t=23$ ms of two different simulations when the light was injected into the active or passive WGs, and indicates that the gold patch does not guide surface plasmon polaritons from one WG to another.
Fig.\ref{NonLocal}f presents the associated phase change in the passive WG as a function of time for three different separations, naturally indicating smaller values of phase shift for increasingly larger separations.
Note, that in contrast to the close separation regime, initially low power (which is not sufficient to deform the liquid film) optical TM mode was injected also into the passive WG in order to analyze its evolution under the effect of the higher power mode in the active WG.}
Notably, the maximal separation between the WGs is much higher than the corresponding evanescent scale and also significantly exceeds other known optical nonlocality scales stemming from alternative optical nonlinearities such as photorefractive \cite{Duree1993}, TO effect \cite{Rotschild2005} and long range molecular interaction between liquid crystals molecules   \cite{McLaughlin1995,Conti2003,Assanto2003}.

\subsection{Self-induced transmittance and reflection in channel Bragg waveguides}

\begin{figure}
	\includegraphics[scale=0.13]{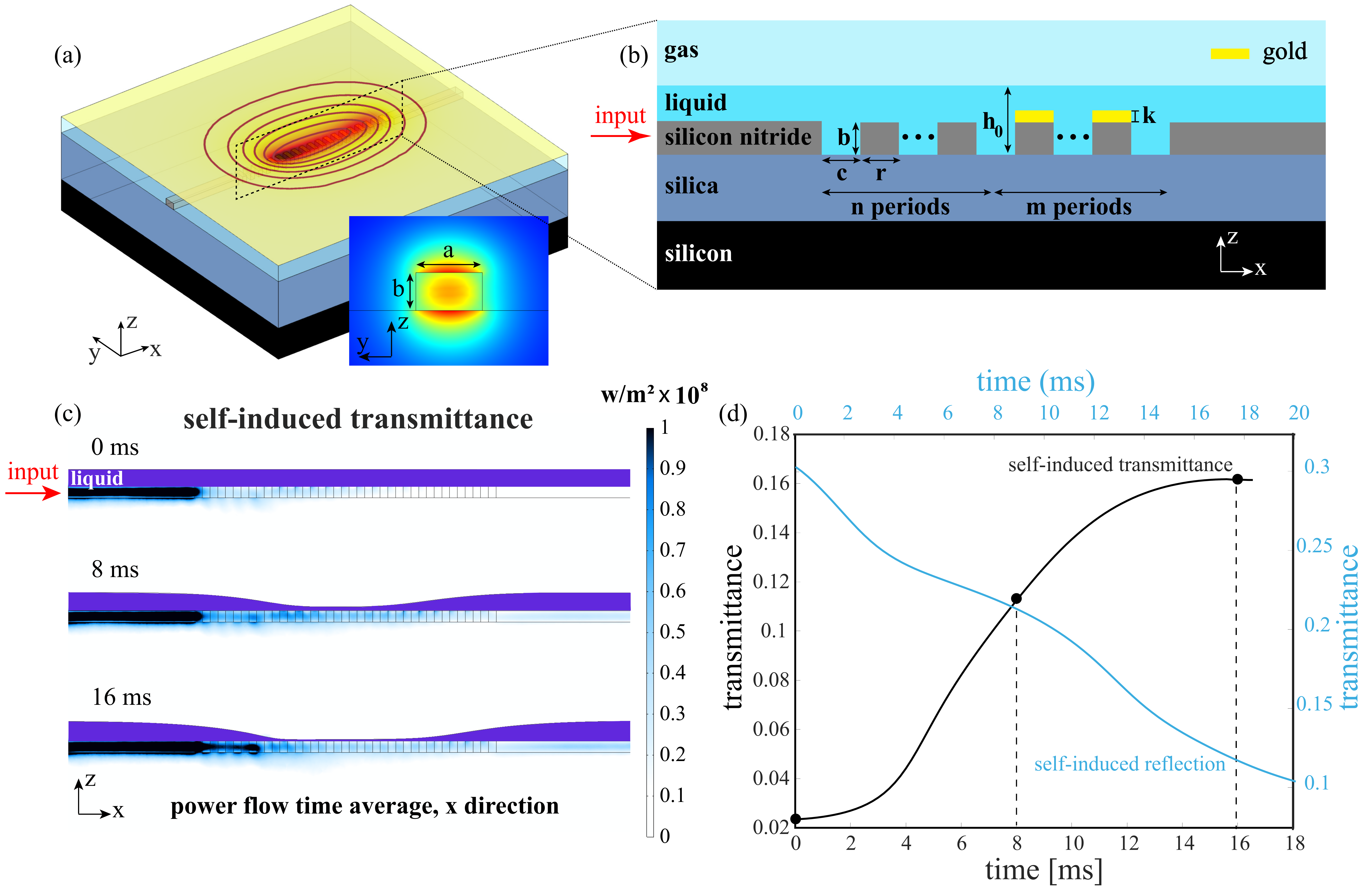} 
    \caption{Numerical simulation results of the self induced transmittance and reflection effects. (a) Presents the underlying 3D geometric setup of the periodic SiN Bragg WG on silica substrate, and (b) the corresponding 2D normal cross section in the $x-z$ plane. Here, $a=0.7$ \text{$\mu$}m, $b=0.4$ \text{$\mu$}m, $c=r=0.2775$ \text{$\mu$}m, $k=20$ nm; 
    number of ribs without gold patch $n=4$ and with gold patch $m=15$ for optimized performance. 
    (c) Colormap of the corresponding power-flow along the WG, as a function of time. 
    At successively later time moments liquid film over the periodic structure becomes increasingly thinner thus facilitating the optical propagation through the periodic structure designed to admit a stop band in the presence of liquid film. 
    (d) Presents transmittance (black curve) as a function of time presenting also the transmittance values for thin liquid topography presented for the corresponding time moments in (c).
    The blue curve presents the decreasing function of transmittance for the complementary case of self-induced reflection where similar structure increasingly rejects light as liquid thickness approaches zero (see SM for more details as well as the video file showing the corresponding 3D liquid deformation as a function of time).}
    \label{SITandSIR}
\end{figure}

Consider SiN Bragg WG on silica substrate presented in Fig.\ref{SITandSIR} where some of the ribs are covered with gold patch of thickness $k$.
Here, we employ SiN rather than Si because the former has a lower refractive index ($2$ at wavelength $1550$ nm) leading to larger mode volume and thus to higher sensitivity of perturbations in the geometry of the gas-liquid interface.
First, we design the periodic WG to admit a stop-band at wavelength $1550$ nm when it is covered with a thin liquid film of thickness $1$ \text{$\mu$}m (as measured from silica substrate), and admit non-zero transmittance without the thin liquid film.
The evanescent penetration of the incident light few periods into the SiN Bragg WG, which is inherent to stop-band conditions, triggers heating of the gold patches close to the input facet (left) which in turn triggers TC-driven deformation of the liquid film. 
As liquid thickness becomes thinner near the input facet, the designed propagation conditions facilitate deeper penetration of the optical mode into the periodic structure thus leading to film thinning above more distant regions from the input region, which in turn amplifies the process.
Indeed, Fig.\ref{SITandSIR}c presents initially flat gas-liquid interface at $t=0$ ms, liquid deformation with indentation closer to the input facet at $t=8$ ms, and wider indentation over the central part of the periodic structure at later time $t=16$ ms leading to almost minimally thin liquid film above all WG ribs.    
Similarly, a periodic structure with slightly different period can be designed to support propagation once the WG is covered with liquid film and stop band once the liquid film is replaced by air.
Following the same arguments as above, the self-induced liquid thinning over the periodic structure is expected to shift the propagation conditions and lead to lower transmittance as indeed described by the blue curve in Fig.\ref{SITandSIR}d.
Interestingly, the presented effects are reminiscent of electromagnetically induced transparency which relies on a very different phenomenon where a quantum interference allows the propagation of light through an otherwise opaque atomic medium \cite{Boyd2008}.

\subsection{Reservoir computing of digital and analog tasks}

\begin{figure}[t]
	\includegraphics[scale=0.32]{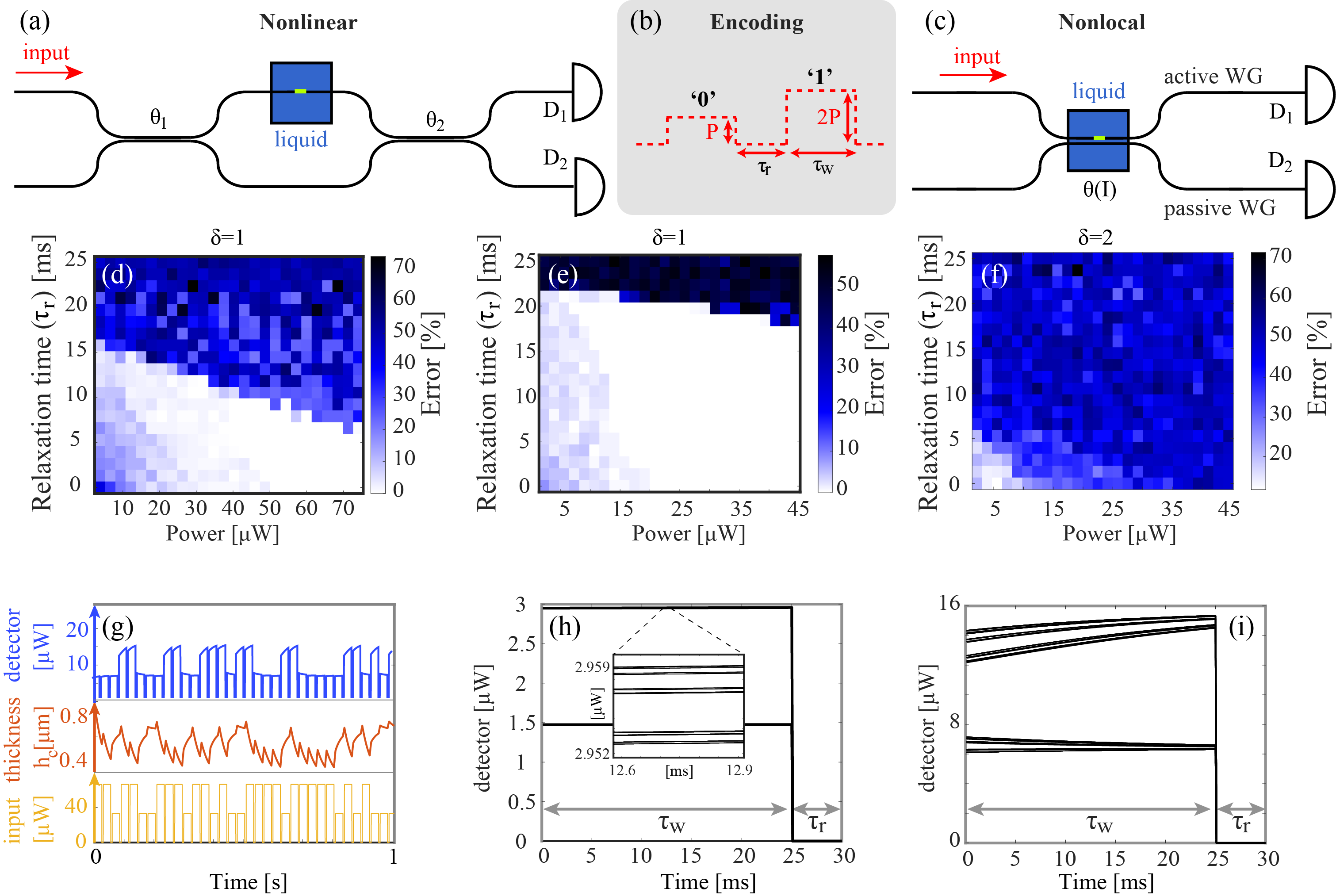}
    \caption{Simulation results presenting RC computing of XOR task by employing self-induced phase change (nonlinear) and self-induced coupling change (nonlocal) effects. 
    (a) Schematic diagram presenting MZI circuit with two linear couplers where the liquid cell introduces self-induced phase change in one of the arms measured by a pair of detectors $D_{1,2}$, 
    (b) encoding scheme of '0' ('1') which employs a square wave modulation with actuation period $\tau_{w}$ of power $2P$ ($P$) followed by relaxation period $\tau_{r}$, and (c) nonlocal circuit which employs self-induced coupling change and identical encoding.
    (d) Test results presenting performance error of XOR task for delay parameter $\delta=1$ as a function of $P_{max}$ and relaxation time $\tau_{r}$ in nonlinear circuit (a), and colormap (e) performing identical task with nonlocal circuit (c) demonstrating enhanced area of vanishing error.  
    (f) Test results of XOR task with $\delta=2$ by using nonlocal circuit (c) showing minimal error $\sim 11 \%$.
    (g) Dynamics evolution of optical signal in detector $D_{1}$ (blue) and liquid thickness $h_{c}$ (red) due to actuation signal (green) with $\tau_{w}=25$ ms, $\tau_{r} = 5$ ms and $P=0.033$ mW. 
    (h) Folded dynamics of optical intensity in nonlinear circuit and (i) nonlocal circuit showing enhanced separability of the different regimes.}
    \label{XOR}
\end{figure}

Consider Fig.\ref{XOR}a  describing Mach-Zehnder interferometer (MZI) circuit with two directional couplers and with an optofluidic cell in one of the arms, which supports the nonlinear self-induced phase change effect described above, and the circuit described by Fig.\ref{XOR}c with a liquid cell placed in the coupling region which affects the transmittance via self-induced coupling change (nonlocal effect).
In both cases, the self-induced phase/coupling change due to liquid deformation invoked by the active WG, are translated into intensity changes in the output arms $D_{1}$ and $D_{2}$.
To allow sufficiently large actuation of the liquid, the corresponding actuation time scale ($\tau_{\ell}$) must be smaller than pulse width ($\tau_{w}$), whereas in order to ensure memory,
i.e., that the liquid deformation imprinted by the $n-1$-th bit will affect the phase/coupling change of the $n$-th bit,
the distance between subsequent pulses ($\tau_{r}$) must be smaller than $\tau_{\ell}$. 
Both conditions can be summarized as
\begin{equation}
    \tau_{r} < \tau_{\ell} < \tau_{w},
\label{TimeScales}    
\end{equation}
and these ensure that the optically imprinted light pulse in gas-liquid interface at time moment $n-1$ affects through the liquid the subsequent pulse at time moment $n$.
Since for an optical signal of given power and operation time, thinner film is expected to introduce more significant nonlinear response (unless it is so thin that the optical signal leads to saturation due to very fast liquid depletion above the gold patch), we choose liquid thickness in the region between $0-0.5$ \text{$\mu$}m.
Below we employ the nonlinear self-induced phase change and the self-induced coupling change (nonlinear effect) between adjacent WGs and the accompanying memory effect, to perform both digital and analog tasks.

\textit{Digital task:} Consider the delayed XOR operation,  where an arbitrary input time series  $\{x_{n}\}_{n=1}^{N}$ ($n=1,2, \cdots, N$) of zeros and ones is mapped to output sequence $\{y_{n}\}_{n=1+\delta}^{N}$ according to
\begin{equation}
    \{x_{n}\}_{n=1}^{N} \longrightarrow \{y_{n}\}_{n=1+\delta}^{N}; \quad y_{n} = x_{n-\delta} \oplus x_{n},
\label{XOReq}    
\end{equation}
where $\oplus$ is the addition modulo two (i.e. XOR operation), $\delta>0$ is an integer encoding the delay. 
In particular, $\delta=1$ corresponds to the simplest case of smallest delay where XOR operation performed on adjacent bits whereas $\delta=2$ corresponds to the case where it is performed on bits which are separated by one bit.
The corresponding dynamics of the thin liquid film, applicable irrespective of nonlocal or nonlinear circuit in Fig.\ref{XOR}, admits the following evolution equation explicitly given by
\begin{equation}
    h_{c}(n) = f \Big[ h_{c}(n-1),P(n) \Big],
\label{RCliquid}
\end{equation}
where the state of the liquid at discrete time moment $n$ is determined by the result of the nonlinear saturation function $f$, which depends on an incident power $P(n)$ at the same time $n$ and liquid thickness at preceding time $n-1$ (see SM for derivation).
This dependence describes memory effect of the liquid film where the finite (typically ms) relaxation time imprinted in gas-liquid interface at time $n-1$ interacts with a subsequent pulse at time $n$.

To implement XOR operation we can either employ the (nonlinear) self-induced phase change or the (nonlinear-nonlocal) 
self-induced coupling change with circuits Fig.\ref{XOR}(a,c), respectively.
The input series $\{ x_{n} \}_{n=1}^{N}$ is encoded as a sequence of square pulses
of power level $P_{n}=P$ or $2P$ depending on the value of logic '0' or '1' as described by Fig.\ref{XOR}b; the corresponding power level operates during a time $\tau_{w}$ and followed by relaxation time $\tau_{r}$.
Fig.\ref{XOR}(d,e) present $P-\tau_{r}$ performance diagram with different colors encoding the corresponding prediction error, for the case of nonlinear circuit and nonlocal circuit described by Fig.\ref{XOR}(a,c), respectively.
Note that performance diagram of nonlocal circuit presents (white) region of vanishing error which has larger area compared to the corresponding area in the performance diagram of nonlinear circuit.
Furthermore, the lowest power value of the white region in the nonlinear case is $\sim 40$ \text{$\mu$}W and lower value around $\sim 10$ \text{$\mu$}W for nonlocal case.
Higher performance of the nonlocal circuit stems from higher sensitivity of the intensity of the transmitted signal with respect to changes of the liquid thickness, compared to the nonlocal case. 
Specifically, sensitivity estimates can be determined by taking the derivative of the MZI signal due to a phase shift $\varphi$, given by $d\cos^{2}(\varphi/2)/dh_{c}=(1/2) \cdot sin(\varphi)d\varphi/dh_{c}$, where the derivative $d\varphi/dh_{c}$ can be estimated from Fig.S\ref{LinearFit}c; similarly Fig.S\ref{CouplingVsIntensity} provides estimate for the typical slope/sensitivity in the nonlocal case.
The enhanced performance of the nonlocal circuit for the delayed $\delta = 1$ XOR task, manifests also for $\delta = 2$ XOR task as can be seen by comparing the corresponding $P-\tau_{r}$ performance diagrams of nonlocal circuit with minimal error $\sim 11 \%$, presented in Fig.\ref{XOR}f, to performance diagram of nonlinear circuit presented in Fig.S\ref{XORd}c with minimal error $\sim 25 \%$. 
In general, as can be seen from the graph Fig.S\ref{XORd}d, 
presenting test error for a fixed $P$ and $\tau_{r}$,
performance of XOR gate quickly degrades for higher values of the delay $\delta$, which also demonstrates prominent memory of only one time backward in time.
The underlying dynamics of the thin liquid film 
was obtained by constructing a reduced 1D model which was obtained from analyzing numerous 3D multiphysics simulations \cite{comsol} results where the thin liquid film was subjected to various initial conditions and actuation powers.
In our reduced model, the transient dynamics of the thin liquid film is subject to first order ordinary differential equation with power dependent coefficients, allowing simple numerical scheme to predict the accompanying phase/coupling change and the resultant power levels at the detectors (see Methods section and SM).
In particular, Fig.S3, Fig.S5a, Fig.S5c in SM present $h_{c}(t)$ as a function of time for several different initial thicknesses and input powers, relation between $\dot{h}_{c}(t)$ and $h_{c}(t)$ and the one-to-one relation between $h_{c}(t)$ and the induced phase shift, respectively, 
whereas Fig.S\ref{CouplingVsIntensityEq} presents transmittance values as a function of optical power in the active WG.
Fig.\ref{XOR}g presents the underlying dynamics of the thin liquid film (red curve) and the output signal (blue curve) in $D_{1}$ detector in nonlocal circuit Fig.\ref{XOR}c, under exciting signal (green curve).
To visually analyze the different patterns of detected power we construct folded dynamics diagrams where optical signal at different time intervals of length $T=\tau_{r}+\tau_{w}$ are plotted on top of each other.
Fig.\ref{XOR}(h,i) present folded dynamics graphs for the nonlinear and nonlocal circuits between times $5T$ to $200T$, respectively, where the signal in the nonlocal circuit shows enhanced separability compared to the nonlinear case performed under similar conditions $\tau_{w}= 25$ ms, $\tau_{r}=5$ ms, and $P=33$ \text{$\mu$}W. 
Interestingly, Fig.\ref{XOR}(h,i) do not show dependence on initial conditions, which is in concert with graph Fig.S\ref{XORd}a demonstrating that the system practically 'forgets' the initial conditions after $5T$, and furthermore showing emergence of patterns which are characteristic of particular dynamics and actuation signals.
For instance, Fig.\ref{XOR}i presents four main curves with internal substructure, corresponding to '1' or '0' which was preceded by '1' or by '0'. 
These represent the four dominant combinations which allow efficient solution of the delayed ($\delta=1$) XOR task, whereas internal structure represents next order memory effects which allows to some extent to solve $\delta=2$ XOR task.  
More formally, since the dynamics of our system converges to states which do not depend on the initial conditions, it satisfies the so-called common-signal-induced synchronization (CSIS) (see \cite{Inubushi} and references within), also known as echo state \cite{Maas2002} which is a necessary property enabling a dynamical system to serve as RC.
To determine the required optical energy $E$ needed for training, we assume that the number of '0' and '1' pulses is equal leading to $E=(NP/2+NP)\tau_{w}$; for instance $N=1000$, $P=10$ \text{$\mu$}W and $\tau_{w}=20$ ms, yields $E=0.1$ J. 
Similarly, the energy during the test stage $E_{test}$, required for solution of a string of length $N_{test}$, is given by $E_{test}=(N_{test}P/2+N_{test}P)\tau_{w}$. 
For the case $N_{test}=20$ and similar parameters used during the training stage discussed above yields $E_{test}=6$ \text{$\mu$}J.

\begin{figure}[h]
	\includegraphics[scale=0.45]{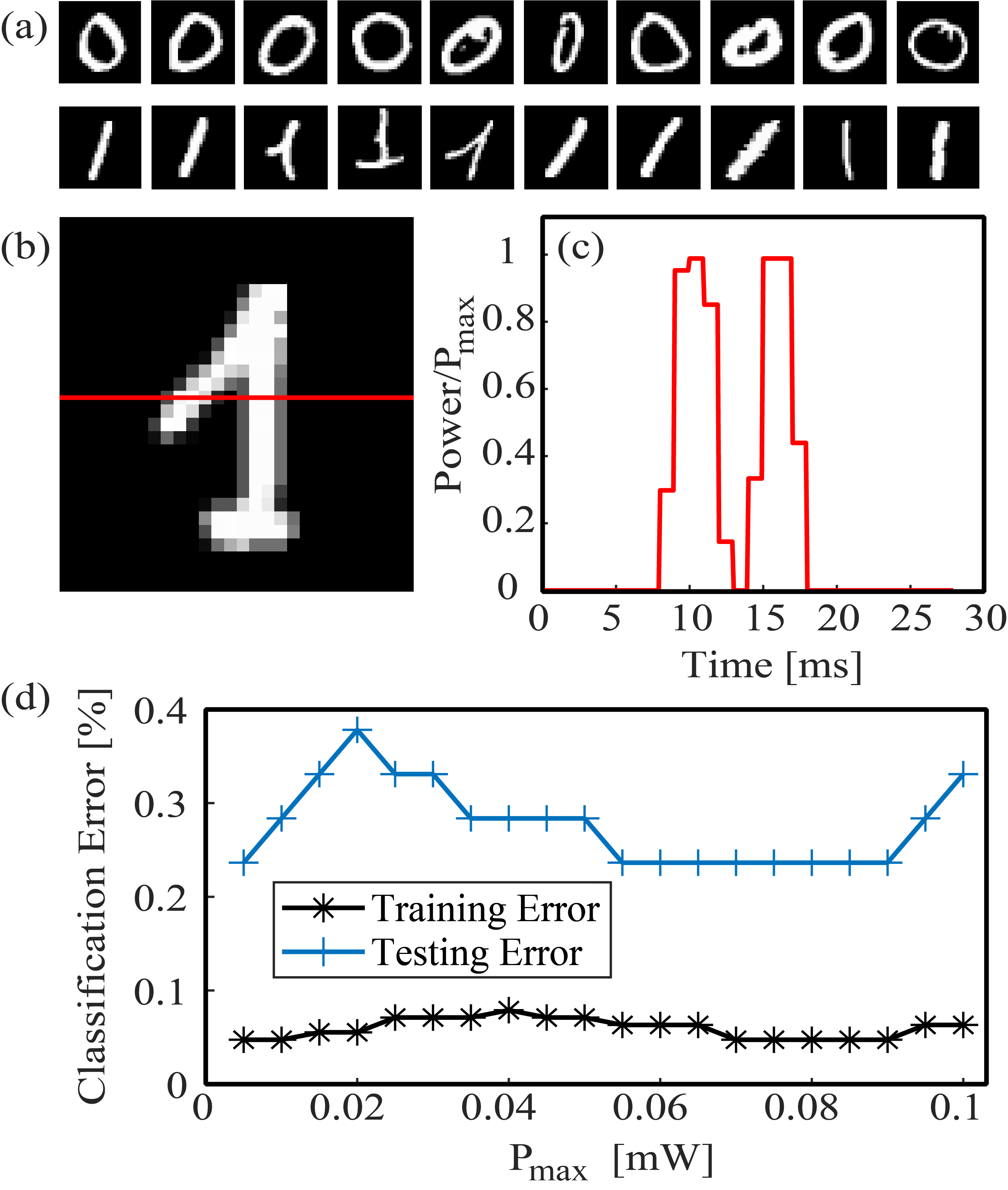}
    \caption{RC of analog task: hand-written 'zeros' and 'ones' classification. (a) Sample of $20$ images each of size $28 \times 28$ pixels, which were used for the classification. 
    (b) Detailed 'one' image with 
    (c) presenting the encoded signal along a single red horizontal row; the bits of two adjacent rows are injected in parallel into the two input arms of nonlocal (Fig.\ref{XOR}a) or nonlinear (Fig.\ref{XOR}c) circuits. 
    (d) Classification error of both training and test stages, each employing $12,665$ and $2,115$ images, respectively. 
    The nonlocal circuit with minimal error $0.14$\% demonstrates enhanced performance compared to nonlinear circuit with minimal error $0.24$\%.
    Employing just one arm of MZI for bits injection, i.e. injecting only one row at a time, increases the classification error for nonlinear circuit to approximately $0.55$\% for similar $P_{max}$ values.}
    \label{01}
\end{figure}

\textit{Analog task:} Next, we consider the classification task of hand-written 'zero' and 'one' digits; Fig.\ref{01}a presents a sample of $20$ images from Modified National Institute of Standard and Technology (MNIST) database \cite{ImagesDatabase}, each of size $28 \times 28$ pixels.
To encode each image as a signal, without filtering or other pre-processing procedure, we partitioned the image into a set of horizontal rows or columns where each pixel is represented by a $1$ ms long constant optical power proportional to the brightness value of the corresponding pixel.
Specifically, the pixels are encoded according to power values $c P_{max}$ where $P_{max}$ is the power needed to encode the brightest pixel of unity brightness ($c=1$), whereas $c=0$ corresponds to dark pixels which carry no features and other values $0<c<1$ correspond to pixels of intermediate brightness.
Fig.\ref{01}b presents an example of a hand-written 'one' digit used for classification where the $28$ ms long signal encoding the highlighted red horizontal row is given by Fig.\ref{01}c with left/right pixels corresponding to early/late times.
To perform image classification we employ 
either nonlinear or nonlocal circuits presented in Fig.\ref{XOR}a and Fig.\ref{XOR}c, respectively,
where the top and the bottom arms receive in parallel the
$1$ ms long bits encoding $i$-th and the following $i+1$-th rows, respectively, thus generating more
correlations between different lines and leading to 
higher accuracy compared to feeding just one arm at a time.
Fig.\ref{01}d presents the classification error during the training and testing stages, with training values naturally admitting lower values. 
Interestingly, both errors admit fairly low values below $0.5$ \% across a wide range of power values $P_{max}$ spanned between $0.01$ mW to $0.1$ mW.
The total energy needed to feed one image is determined by $28 \times 28 \times 1 \text{ ms} \times P_{max} \times \bar{c}$, where $\bar{c}$ is the average brightness level of the image laying between $0<\bar{c}<1$ and typically admitting value $\sim 0.15$. 
The corresponding energy for training/testing stage is achieved by multiplying the one image energy above, by the total number of corresponding images (see Methods section); for $P_{max}=10$ \text{$\mu$}W the corresponding average power level is given by $\bar{c}P_{max}=1.5$ \text{$\mu$}W.
Importantly, the single liquid cell circuits in Fig.\ref{XOR}(a,c) can be extended to form more complex network. 
For instance, Fig.S\ref{Network} presents a network consisting of four inputs, two liquid cells and two detectors. 
While it does not present enhancement in accuracy it allows to reduce the computation time by a factor of two without increasing the total required energy. 
Analyzing networks with more inputs, allowing to achieve optimization of handwritten digits recognition task and further reduction in operation time is beyond the scope of this work.

To learn more about the role of the reservoir in our modality, we also performed linear regression analysis where the output signal is made equal to the input signal which is followed by ridge regression allowing to obtain the corresponding weights that perform the corresponding task with corresponding error given by $0.61\%$.
Interestingly, employing a reservoir presented in Fig.\ref{XOR}a but without liquid response, yields similar $0.61\%$ performance for ‘zero’/’one’ digits classification, which is around $4.3$ times higher compared to the highest $0.14\%$ reservoir performance for this task.
It is worth mentioning that the performance described in Fig.\ref{XOR}d and Fig.S\ref{01_columns} is not a prominent function of the optical power and using larger power doesn’t lead to significantly enhanced performance.
While in this work the input data used for the analog task was not subject to any preprocessing, we expect that common methods such as edge detection should improve the accuracy.

\section*{Discussion}


To summarize, we believe that the family of physical effects predicted to take place due to nonlinear-nonlocal interaction between thin liquid film to channel WG modes in integrated chip-scale photonics platforms, and the emergent concept suggesting to employ these effects to realize optofluidic RNN supporting RC applications, is expected to stimulate both pure light-matter interaction studies and development of novel RNN architectures/schemes.
In particular, our 3D simulation results which take into account optical propagation with losses, heat transport, fluid dynamics as well as the underlying surface tension physics, 
indicate that the magnitude of the self-induced phase change due to the TC effect in the liquid film, is approximately three orders of magnitude higher compared to the more common heat-based TO effect. 
Therefore, the predicted effects of self-induced coupling change and self-induced bandstructure and transmittance change can stimulate experimental and theoretical studies validating the predictions and exploring
bandstructure tuning in photonic crystals and its topological properties \cite{Lu2014,Ozawa2019}, as well as
studying nonlocal interaction between large number of WGs even if their separation exceeds the optical evanescent scales.
\textcolor{black}{While in our work we employed gold patch due to its optical absorption properties, studying metal structures designed to support resonant absorption due to surface plasmon polariton oscillations is another intriguing possibility capable to induce and modify resonant absorption in channel WGs as we briefly mention in Fig.S11.}
From ML-based perspective, the employed
self-induced phase/coupling change and allowing to achieve nonlocal-nonlinear interaction and memory at the same compact actuation region, and capable to perform digital XOR and analog handwritten images classification task, can lead to new RNN elements and architecture which take advantage of the built-in internal feedback capability.
Notably, our work demonstrates for the first time that controlled deformation of gas-liquid interface on sub-micron scale, can operate as RNN which can be extended to a network with larger number of liquid cells (or larger number of WGs in a single liquid cell), and further provides design principles of compact optofluidic and complementary metal-oxide-semiconductor (CMOS) compatible optofluidic RC systems 
which are capable to induce additional nonlinearity beyond the commonly employed square-law detection, without utilizing optical resonant structures, and furthermore operates under wide range of conditions which do not require phase matching conditions.
In particular, our design of optofluidic-based RC system is capable to lead to four-five orders of magnitude reduction in size compared to previous liquid-based RC \cite{Fernando2003, Adamatzky2002} systems, and also to few orders of magnitude faster computation compared to the surface waves dynamics and reaction-diffusion processes which were employed in \cite{Fernando2003} and \cite{Adamatzky2002}, respectively.
While the presented optofluidic approach for RC requires more energy compared to the present state-of-the-art approaches such as opto-electronic based method, offering $\sim 10$ GHz modulation rate with hundreds of fJ per bit \cite{Sun2019} (see Table S2), we should keep in mind that Electro-optic approach requires external modulation unit whereas in Optofluidic approach the modulation is internal due to the specificity of the gas-liquid interaction. 
In particular, internal feedback allows to invoke nonlinearities and memory effects in the interaction region and thus translates into very compact computation region without feedback loops which are required in systems that employ external feedback.
It is not clear how to achieve self-induced refractive index response of order O$(10^{-1})$ with Electro-optic effect, though self-induced modulation can be achieved by employing two photon absorption and other nonlinearities \cite{Mesaritakis2013} by using significantly higher power levels compared to state-of-the-art externally modulated mirco-ring resonator devices \cite{Sun2019}. 
Furthermore, we should keep in mind that neural brain activity also occurs on a ms time scale and is based on several ionic transport processes which also take place in liquid environment. 
Therefore, even if our study in its current version cannot suggest computational capabilities comparable to that of a human brain, the general strategy of combining several different physical processes rather than relying on a small number of physical effects, may prove to be as a useful strategy to achieve more efficient computational approaches.      
It is worth mentioning that while phase-change materials, where optically induced phase change between crystalline and amorphous phases invokes self-induced transmittance change, were recently employed for NC applications \cite{Feldman2019,Rios2015,Wuttig2017}, 
\textcolor{black}{the corresponding material phase change is not able to nonlocally affect nearby WGs.}
Given the fact that in this work we demonstrated that nonlocality enhances performance of both digital and analog tasks, we believe that our study will motivate future studies with more efficient optofluidic-based architectures which employ nonlocality effects, and in parallel also stimulate material science-oriented studies how nonlocality can be harnessed in more conventional phase-change materials.
Noteworthy, we performed preparatory experiments demonstrating controlled deposition of non-volatile liquid into etched trenches above silicon channel WG, bringing future experimental realization of the predicted effects within a closer reach (see Fig.S\ref{Deposition}). 

\section*{Acknowledgements}

The authors cordially thank Defense Advanced Research Projects Agency (DARPA), Nature as Computer (NAC) program management team for stimulating discussions during the program. 
This work was supported by the Defense Advanced Research Projects Agency (DARPA) DSO's NAC (HR00112090009) and NLM Programs, the Office of Naval Research (ONR), the National Science Foundation (NSF) grants CBET-1704085, DMR-1707641, NSF ECCS-180789, NSF ECCS-190184, NSF ECCS-2023730, the Army Research Office (ARO), the San Diego Nanotechnology Infrastructure (SDNI) supported by the NSF National Nanotechnology Coordinated Infrastructure (grant ECCS-2025752), the Quantum Materials for Energy Efficient Neuromorphic Computing - an Energy Frontier Research Center funded by the U.S. Department of Energy (DOE) Office of Science, Basic Energy Sciences under award \#DE-SC0019273, and the Cymer Corporation.


\widetext
\begin{center}
\newpage
\title{SI}
\textbf{SUPPLEMENTAL MATERIAL for: \\
Thin liquid film as an optical nonlinear-nonlocal media and memory element in integrated optofluidic reservoir computer}

\text{Chengkuan Gao, Prabhav Gaur, Shimon Rubin and Yeshaiahu Fainman}

\textit{Department of Electrical and Computer Engineering, University of California, San Diego, 9500 Gilman Dr., La Jolla, California 92023, USA}
\end{center}

\setcounter{equation}{0}
\setcounter{figure}{0}
\setcounter{section}{0}
\setcounter{table}{0}
\setcounter{page}{1}

\renewcommand{\thesection}{S.\arabic{section}}
\renewcommand{\thesubsection}{\thesection.\arabic{subsection}}
\makeatletter 
\def\tagform@#1{\maketag@@@{(S\ignorespaces#1\unskip\@@italiccorr)}}
\makeatother
\makeatletter
\makeatletter \renewcommand{\fnum@figure}
{\figurename~S\thefigure}
\makeatother
\makeatletter \renewcommand{\fnum@table}
{\tablename~S\thetable}
\makeatother

\section{Governing equations and key parameters}


Below we present the governing equations for the optical field, heat transport, fluid dynamics and the matching conditions across the gas-liquid interface. The coupling scheme between the four mechanisms operates as follows. 
First, the propagating optical WG mode interacts with the gold patch and induces heating which is triggers temperature gradient of the gas-liquid interface, leading in turn to gradients of the surface tension. 
The latter invokes TC flows and thickness changes of the liquid film, leading to changes of the effective refractive index of the optical mode. 

\subsection{Optical field}

Faraday and Ampere's law equations relating between electrical field $\vec{E}$, magnetic field $\vec{B}$, electric displacement $\vec{D}$, magnetic induction $\vec{H}$ and current density $\vec{J}$ are given by
\begin{subequations}
\begin{align}
    \vec{\nabla} \times \vec{E} &= -\dfrac{\partial \vec{B}}{\partial t}
\\
    \vec{\nabla} \times \vec{H} &= \dfrac{\partial \vec{D}}{\partial t} + \vec{J}.
\end{align}
\label{Maxwell}
\end{subequations}
Applying $\vec{\nabla} \times$ operator on Eq.S\ref{Maxwell}a and then employing: constitutive relation $\vec{B} = \mu \vec{H}$, Eq.S\ref{Maxwell}b and Joule's law $\vec{J} = \sigma \vec{E}$, yields 
\begin{equation}
    \vec{\nabla} \times \left(\dfrac{1}{\mu_{r}} \vec{\nabla} \times \vec{E} \right) = k_{0}^{2} \left( \epsilon_{r} - \dfrac{i \sigma}{\omega \epsilon_{0}} \right) \vec{E}.
\end{equation}
Here, $k_{0}^{2} = \omega^{2} \epsilon_{0};  \epsilon_{0}$ and $\mu_{0}$ are electric and magnetic vacuum permittivities, respectively; $\epsilon_{r}$ and $\mu_{r}$ are electric permittivity and magnetic permittivity, respectively. 

\subsection{Heat transport}

Heat diffusion transport is governed by the following diffusion equation for the temperature field $T$, 
\begin{equation}
    \rho_{m} c_{p} \dfrac{\partial T}{\partial t} - \vec{\nabla} \cdot \left( k_{th} \vec{\nabla} T \right)  = \vec{J} \cdot \vec{E},
\end{equation}
Where $\rho_{m}$, $c_{p}$, $k_{th}$ are the corresponding material density, specific heat and heat conductivity, and the term $\vec{J} \cdot \vec{E}$ is the Joule heat source term.

\subsection{Fluid dynamics of liquid and gas}

Dynamics of a Newtonian, non-compressible fluid in cartesian coordinates is governed by the Navier-Stokes equations given by
\begin{equation}
    \rho \left( \dfrac{\partial u_{i}}{\partial t} +  u_{j} \dfrac{\partial}{\partial x^{j}} u_{i} \right) =\dfrac{\partial}{\partial x^{j}} \tau_{ij} + F_{i}; \quad i,j=x,y,z,
\end{equation}
where $\tau_{ij}$ is the corresponding energy-momentum tensor given by
\begin{equation}
    \tau_{ij} = -p\delta_{ij} + \mu e_{ij}; \quad e_{ij} = \dfrac{\partial u_{i}}{\partial x_{j}} + \dfrac{\partial u_{j}}{\partial x_{i}} - \dfrac{2}{3} \delta_{ij}\dfrac{\partial u_{l}}{\partial x_{l}}.
\end{equation}
Here, $u_{i}$, $F_{i}$, $\rho$, $p$ and $\mu$ are the fluid velocity components, body force components, density, pressure, and  dynamic  viscosity,  respectively. The indices $i,j$ run  over  the three  Cartesian  coordinates $x,y,z$,  and  summation  convention  over  repeated  indices is employed.

\subsection{Gas-liquid matching conditions}

The interfacial Stress Balance Equation (SBE) which holds on the gas-liquid (or liquid-liquid or gas-gas) interface, is given by the following matching conditions
\begin{equation}
    \hat{n}_{j} \cdot \Big[ \tau_{ij}^{(2)} - \tau_{ij}^{(1)} \Big] = \sigma \hat{n}_{i} \dfrac{\partial \hat{n}_{j}}{\partial x_{j}} - \dfrac{\partial \sigma}{\partial x_{i}},
\end{equation}
which in vector notation takes the following form
\begin{equation}
    \hat{n} \cdot \left( -p I + \mu \left( \vec{\nabla} \vec{u}^{T} + \vec{\nabla} \vec{u} - \dfrac{2}{3} \vec{\nabla} \cdot \vec{u} \right) \right) = \sigma \hat{n} (\vec{\nabla} \cdot \hat{n}) - \vec{\nabla} \sigma.
\end{equation}
Normal stress balance and tangential stress balance are obtained by projecting SBE on $\hat{n}$ and $\hat{t}$, respectively, and are given by
\begin{subequations}
\begin{align}
    \hat{n} \cdot \left( -p I + \mu \left( \vec{\nabla} \vec{u}^{T} + \vec{\nabla} \vec{u} - \dfrac{2}{3} \vec{\nabla} \cdot \vec{u}\right) \right) \cdot \hat{n} =
    \sigma (\vec{\nabla} \cdot \hat{n}),
\\    
       \hat{n} \cdot \left( -p I + \mu \left( \vec{\nabla} \vec{u}^{T} + \vec{\nabla} \vec{u} - \dfrac{2}{3} \vec{\nabla} \cdot \vec{u} \right) \right) \cdot \hat{t} =
    \sigma_{T} \nabla_{t} T,
\end{align}
\label{projections}
\end{subequations}
where in the last line we used $ \hat{t} \cdot \vec{\nabla} \sigma \equiv \nabla_{t} \sigma$ and assumed the commonly employed linear dependence of surface tension on temperature $\sigma(T) = \sigma_{0} - \sigma_{T} T$ ($\sigma_{0}$ and $\sigma_{T}$ are typically positive constants).

\subsection{Boundary conditions}

Heat transport: Dirichlet boundary conditions of fixed temperature $20^{o}$ on the boundary.

Liquid: Navier boundary conditions on the vertical walls with built-in factor of minimum element length equal to one. On horizontal walls we employed vanishing slip velocity conditions.

Optical field: PML boundary conditions in order to reduce reflections from the boundaries.

\subsection{Numerical values used}

The table below specifies the numerical values employed in the multiphysics simulations. All parameters are at $293.15$ K, $1$ atm, and for wavelength $1550$ nm. 

\begin{table}[h]
\centering
\begin{tabular}{ |p{7cm}||p{1.4cm}||p{1cm}||p{1cm}||p{1cm}||p{1.3cm}||p{1cm}|  }
 \hline
   & gas & liquid  & Si & Si$0_{2}$ & Gold  &  SiN \\
 \hline
 Thermal conductivity, $k_{th}$ [W/(m·K)]   & 0.026 & 0.15 & 130 & 1.38 & 310 & 30  \\
  \hline
 Density, $\rho$ [kg/m$^{3}$] & 1.2	& 930 & 2329 & 2203 & 19300 &	2500  \\
  \hline
 Constant pressure heat capacity, $c_{p}$ [J/(kg·K)] & 1005.5	& 1500	&700	&703	&125	&170  \\
  \hline
 complex refractive index, n + i·k &1	&1.444	&3.4757	&1.444	&0.52406 + 10.742·i	&2  \\
  \hline
 Dynamic viscosity, $\mu$ [Pa·s] & 1.8·10$^{-5}$ & 0.1 & - & - & - & -  \\
  \hline
 Ratio of specific heats, $c_{p}/c_{V}$ & 1.4 & 1.5 & - & - & - & -  \\
 \hline
\end{tabular}
\caption{Key physical parameters employed in the multiphysics simulation.}
\end{table}

Liquid's surface tension is assumed to depend linearly on the temperature, $\sigma(T) = \sigma_{0} - \sigma_{T} (T - T_{0})$, where $T_{0}$ is the ambient temperature,  $\sigma_{0} =10^{-3}$ N/m, and $\sigma_{T} = 10^{-4}$ N/(m·K). 

\section{Comparison between TC-based and TO-based self-induced phase change}

\begin{figure}[ht]
\centering
\includegraphics[scale=0.6]{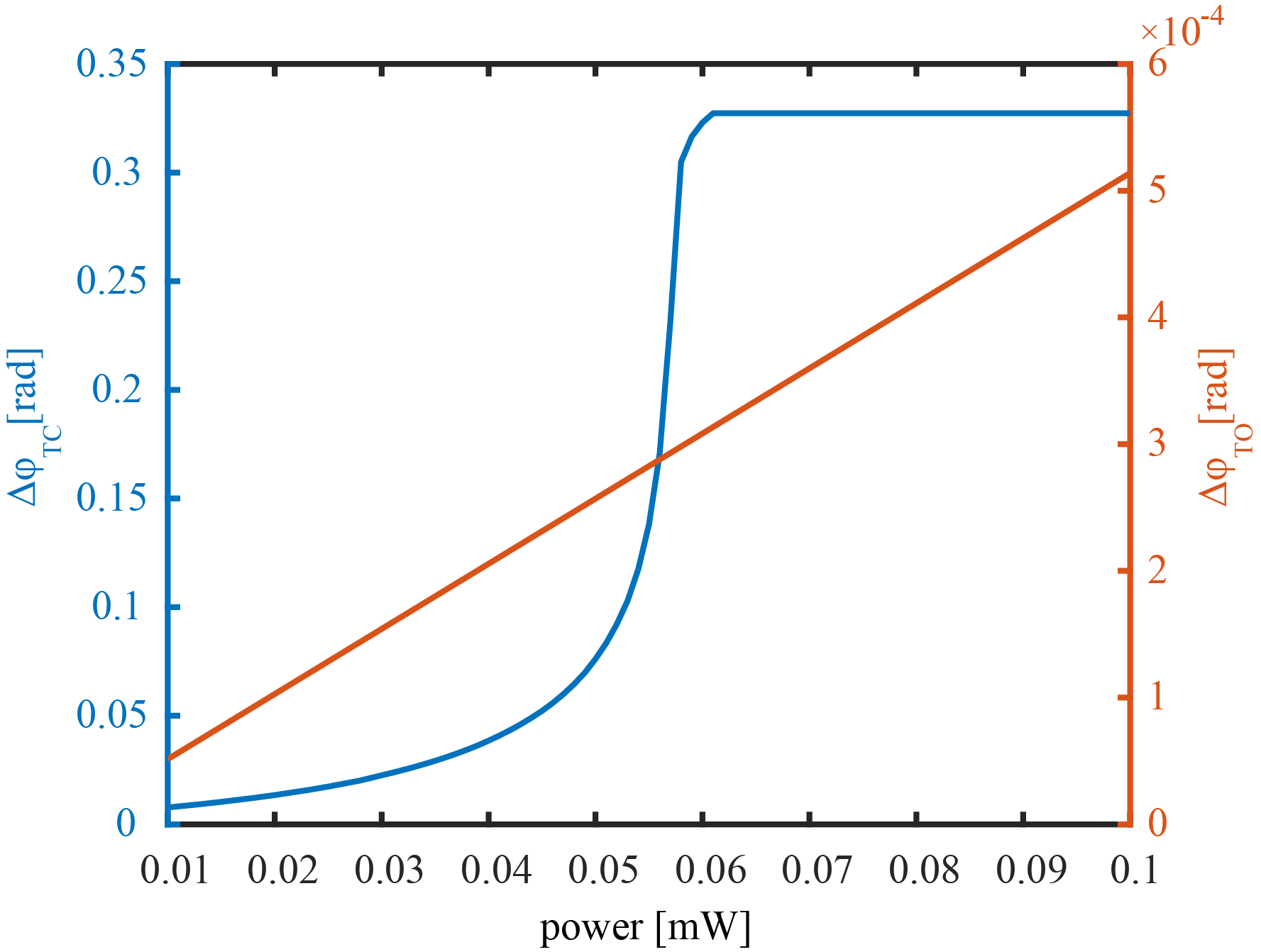}
         \caption{Simulation results comparing the self-induced phase changes $\Delta \varphi_{TC}$ and $\Delta \varphi_{TO}$ under TC and TO effects, respectively.
         The result indicates that TC-driven mechanism is approximately three orders of magnitude larger compared to TO effect under similar optical powers. The value of TO coefficient of silicon employed in our simulation is $dn/dT = 1.8 \cdot 10^{-4}$ 1/K [43].}
\label{TOTC}
\end{figure}

To compare the self-induced phase change due to TC and TO effects, denoted by $\Delta \varphi_{TC}$ and $\Delta \varphi_{TO}$, respectively, we compare simulation results of a single active WG described in Fig.2 to a similar simulation where the $500$ nm thick liquid film is replaced by a solid film of identical thickness and refractive index, but which is not allowed to deform.
Fig.S\ref{TOTC} presents comparison between the self-induced phase change of the two cases indicating that the magnitude of the effect is of the order $\Delta \varphi_{TC}/\Delta \varphi_{TO} \sim 500-1000$
Furthermore, while the $\Delta \varphi_{TO}$ presents linear behavior in optical power, $\Delta \varphi_{TO}$ presents nonlinear behavior due to nonlinear relation and also to saturation due to depletion of the liquid above the gold patch.

\section{Self-induced reflection}

\begin{figure}[ht]
\centering
\includegraphics[scale=0.15]{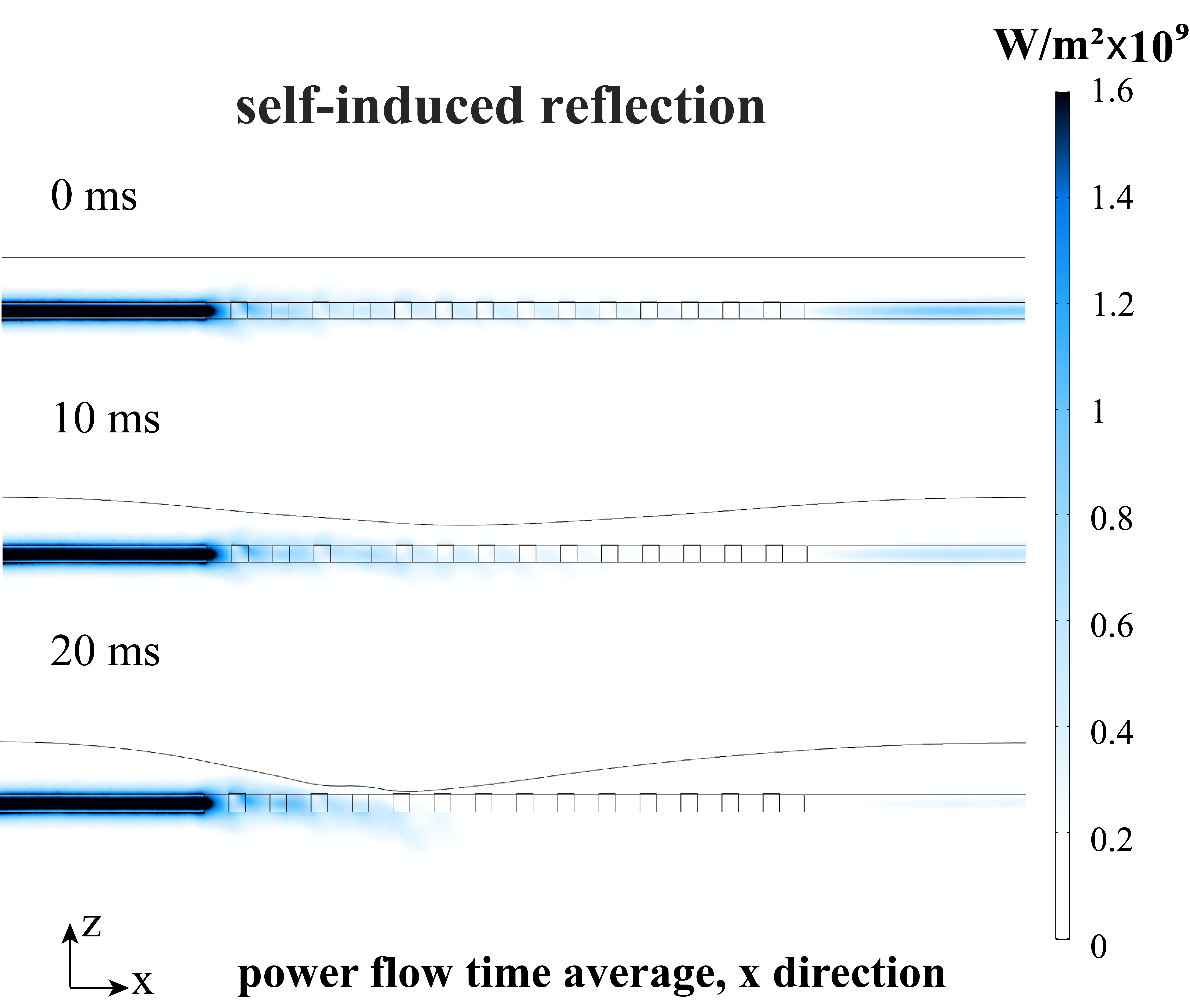}
         \caption{Numerical results presenting the evolving geometry of the gas-liquid interface during the process of self-induced reflection, at time moments $0$ ms, $10$ ms, $20$ ms, thus complementing the results presented in Fig.4.}
\label{SIR}
\end{figure}

Fig.S\ref{SIR} presents 3D simulation results
of the gas-liquid interface evolving geometry  during the process of self-induced reflection,
at time moments $0$ ms, $10$ ms, $20$ ms.
The transmittance during this process is described by the blue curve in Fig.4d, indicating lower transmittance values as liquid film above the Bragg WG becomes thinner.
 Similarly to the case of self-induced transmittance, Fig.4c presents the geometry of the Bragg WG designed to support the case of self-induced reflection with the following parameters: $a=0.7$ \text{$\mu$}m, $b=0.4$ \text{$\mu$}m, $c=0.6$ \text{$\mu$}m, $d=0.4$ \text{$\mu$}m, $k=0.02$ \text{$\mu$}m, and gold patch on all $15$ SiN ribs except the second and the fourth rib.

\section{Evolution of dynamical process as RNN and proof of Eq.3}

First order partial differential equation governing evolution of physical observable $A$ can be written as 
\begin{equation}
    \dfrac{\partial A(\vec{r},t)}{\partial t} + H\big[ A(\vec{r},t),\vec{\nabla}A(\vec{r},t),t \big] A(\vec{r},P(t),t) = J(t),
\label{FirstPDE}    
\end{equation}
where $H$ is a function which depends on $A$, its spatial derivatives $\vec{\nabla}A$, and time dependent coefficient $P(t)$; $J(t)$ is an additional source term. 
Introducing spatial discretization and grid points labelled by vectors $\alpha,\beta$ (e.g. $\alpha=(x_{\alpha},y_{\alpha},z_{\alpha})$ is a position vector pointing to the relevant grid point in the 3D space), 
allows to rewrite Eq.S\ref{FirstPDE} as the following system of first order ordinary differential equations
\begin{equation}
    \dfrac{dA_{\alpha}(t)}{d t} + H_{\alpha \beta}\big[ A_{\gamma}(t), P(t) \big] A_{\beta}(t) = J_{\alpha}(t),   
\label{FirstODE}    
\end{equation}
where $A_{\alpha}(t) \equiv A(\vec{r}_{\alpha},t)$, $J_{\alpha}(t) \equiv J(\vec{r}_{\alpha},t)$ and $H_{\alpha \beta} \equiv H(\vec{r}_{\alpha},\vec{r}_{\beta},t)$ are the values of the corresponding quantities sampled at the relevant spatial points, summation over repeated indices is assumed, and $\alpha$ runs over all grid points.
Employing time discretization with small time step $\Delta t$ (for the particular case of piece-wise constant excitation described in Fig.5 the time step coincides with $\tau_{W,r}$ which isn't necessarily small), allows to rewrite Eq.S\ref{FirstODE} as 
\begin{equation}
    A_{\alpha}(t) = A_{\alpha}(t-\Delta t) - \Delta t \cdot \left( H_{\alpha \beta}\big[ A_{\gamma}(t),P(t) \big] A_{\beta}(t) + J_{\alpha}(t) \right), 
\label{Aalpha}    
\end{equation}
where we have used the Calculus theorem stating that if derivative exists then its left and right limits exist and are equal.
Substituting the expression for $A_{\alpha}(t)$ given by Eq.S\ref{Aalpha} to express $A_{\beta}(t), A_{\gamma}(t)$ in the right hand side of Eq.S\ref{Aalpha}, and furthermore keeping first order terms in $\Delta t$ yields
\begin{equation}
\begin{split}
    &A_{\alpha}(t) = f \big[ A_{\alpha}(t-\Delta t),P(t) \big] \equiv 
\\    
    A_{\alpha}(t-\Delta t) - &\Delta t \cdot \left( H_{\alpha \beta}\big[ A_{\gamma}(t - \Delta t),P(t) \big] A_{\beta}(t - \Delta t) + J_{\alpha}(t) \right), 
\label{AalphaF}
\end{split}
\end{equation}
which admits a functional form of RNN update equation similar to Eq.3.
While Eq.S\ref{FirstPDE} describes dynamics of various degrees of freedom such as heat transport, deformation of thin liquid films, mass transport and quantum wave-function, the approach above can be generalized to practically any system including those described by second order systems in time.
For the particular case of thin liquid film of constant viscosity, vanishing body and surface forces, as well as vanishing slip velocity on the boundary, Eq.S\ref{FirstPDE} takes the form
\begin{equation}
    \dfrac{\partial h}{\partial t} + \dfrac{1}{\mu} \vec{\nabla}_{\parallel} \cdot \Big(  \dfrac{1}{2} h^{2} \vec{\nabla}_{\parallel} \sigma + \dfrac{\sigma_{0}}{3} h^{3} \vec{\nabla}_{\parallel}^{ } \nabla_{\parallel}^{2}h  \Big) = 0,
\end{equation}
which can be formally rewritten as Eq.S\ref{FirstODE}
\begin{equation}
    \dfrac{\partial h}{\partial t} + H(h) h =0; \quad 
    H(h) \equiv \dfrac{1}{\mu} \left( 
    \dfrac{h}{2} \nabla_{\parallel}^{2} \sigma + 
   \vec{\nabla}_{\parallel} h \cdot \vec{\nabla}_{\parallel} \sigma + 
    \dfrac{\sigma_{0} h^{2}}{3} \nabla_{\parallel}^{4}h +  
    \sigma_{0} h \vec{\nabla}_{\parallel} h \cdot \vec{\nabla}_{\parallel} \nabla_{\parallel}^{2} h
    \right).
\label{EvolutionTLD}    
\end{equation}
Consequnetly, following the same steps as above one can readily rewrite Eq.S\ref{EvolutionTLD} as RNN update equation Eq.S\ref{AalphaF} (i.e. Eq.3 in the main text with different notation for the time argument).



\section{Constructing RC simulation}

To construct RC simulation we employed the following three steps: (i) collected dynamics of $h_{c}(t)$ as a function of time for numerous driving optical powers and initial conditions; 
(ii) constructed reduced 1D model describing dynamics of $h_{c}(t)$ under first order ordinary different equation with power dependent coefficients; 
(iii) employed the reduced evolution equation to simulate dynamics of thin liquid film under driving optical sequence, collect the corresponding output data and then perform reservoir training and test its performance.

\subsection{Dynamics of $h_{c}(t)$ as a function of time for numerous driving optical powers and initial conditions}

\begin{figure}[ht]
\centering
\includegraphics[scale=0.48]{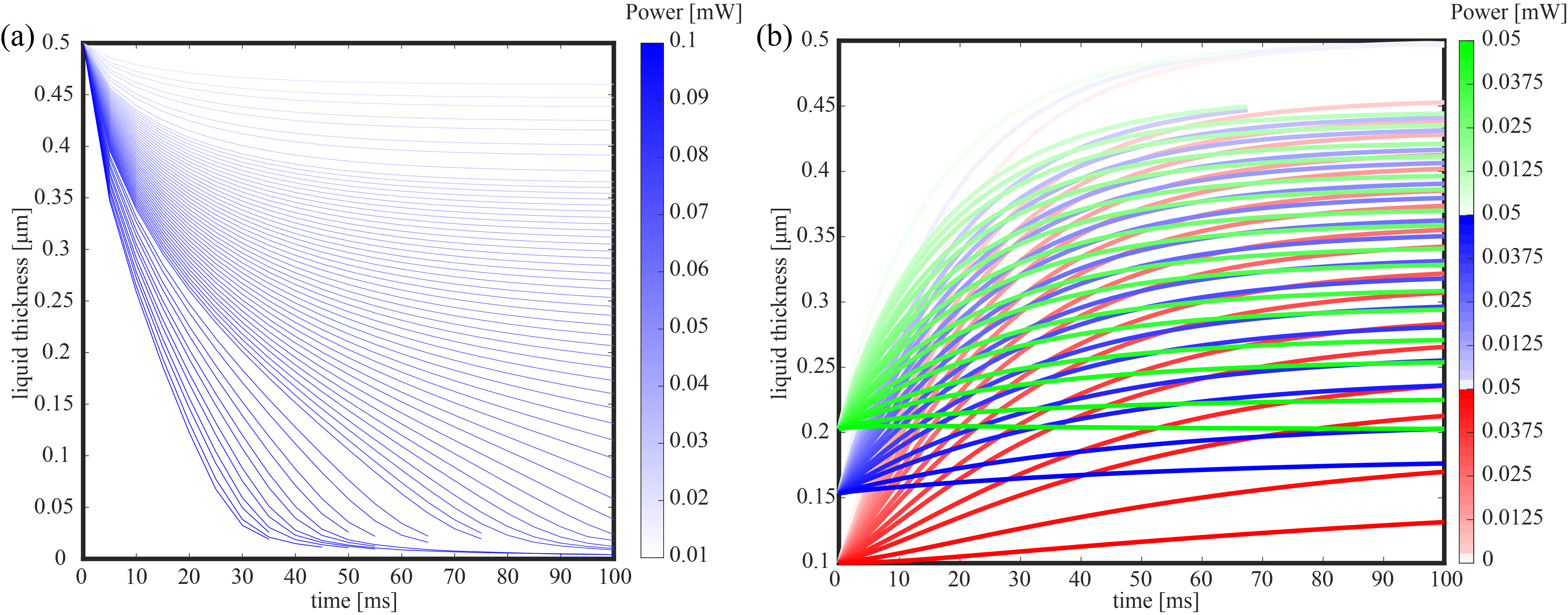}
         \caption{Numerical results presenting thickness of thin liquid film of initial thickness $h_{c}(0) = 500$ nm at a central point above the gold patch as a function of time for various optical powers and various initial conditions. 
         (a) Liquid film thickness as a function of time under approximately $65$ different power levels in the range bracketed by $0.01$ mW and $0.1$ mW describing film thinning. 
         (b) Liquid thickness $h_{c}(t)$ for initial conditions $0.1, 0.15, 0.2$ \text{$\mu$}m; each liquid thickness of corresponding to the initial conditions above is subject under approximately $15$ different power levels in the range below $0.05$ mW corresponding to film relaxation to thicker configuration.}
\label{curves}
\end{figure}

\begin{figure}[t]
\centering
\includegraphics[scale=0.30]{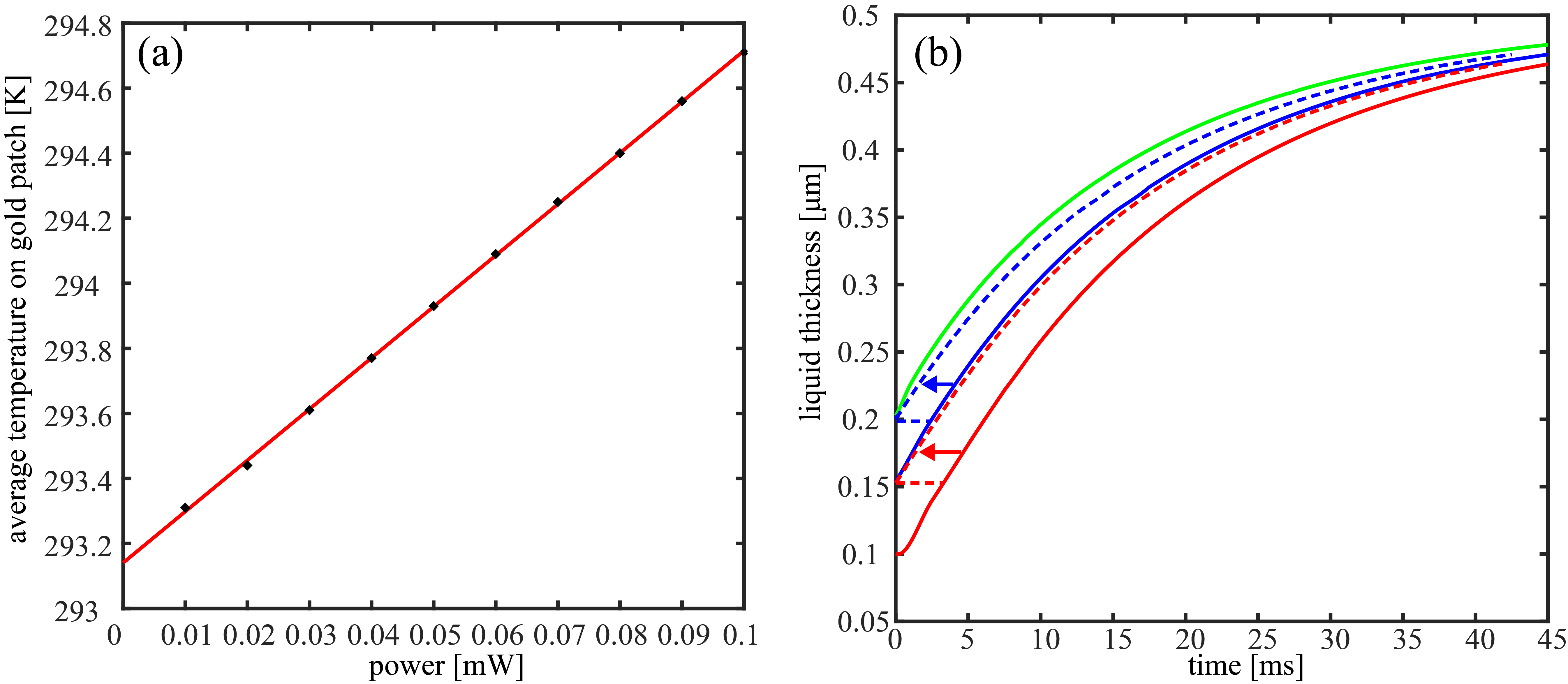}
         \caption{Simulation results summarizing the simplifying assumption followed during the construction of RC simulation. 
         (a) Linear relation between the optical power ($P$) in the WG to the mean temperature ($T$) in the gold patch on top of the WG enabling to replace the optical source by equivalent effective temperature boundary condition. 
         (b) Comparison of relaxation dynamics under different initial conditions, enabling to implement a single model for all cases irrespective of previous history.}
\label{P_Temp}
\end{figure}

 Fig.S\ref{curves} presents evolution of $h_{c}(t)$ as a function of various optical power levels and several initial conditions which reflect the anticipated working regime.
 The optical power levels used to generate the curves described by Fig.S\ref{curves}a in the following ranges are: in range $0.01 - 0.03$ mW with alternating steps $0.003$ mW and $0.002$ mW; in range $0.031-0.07$ mW with constant step of $0.001$ mW; in range $0.073-0.1$ mW with alternating steps $0.003$ mW and $0.002$ mW. 
 For restoring curves presented in Fig.S\ref{curves}b we employed powers between $0.01 - 0.05$ mW with alternating steps of $0.003$ and $0.002$ mW.
 Importantly, to generate the curves we employed temperature equivalent optical power allowing to significantly reduce the computational cost of each model in terms of memory and computation time at the cost of introducing a  small error.
Since under optical excitation the temperature reaches equilibrium on time scale of few milliseconds, which is sufficiently smaller compared to time scales governing evolution of thin liquid film, introducing equivalent temperature source won't affect significantly the liquid dynamics during the relevant time scale.
Furthermore, since our RC simulation operates with liquid films of thickness above $100$ nm, we expect that changes of the optical mode due to evolving liquid thickness, will not lead to significant changes of optical heat dissipation and temperature values. 
Fig.S\ref{P_Temp} summarizes some of the simplifying assumptions we employed during the construction of the reservoir simulation.
Fig.S\ref{P_Temp}a presents the mean steady state temperature ($T$) in the gold patch for $10$ cases of different optical power level ($P$), and the corresponding linear fit which is explicitly given by
\begin{equation}
    T = 15.75 \cdot P + 293.15. 
\end{equation}
Fig.S\ref{P_Temp}b presents three evolution curves of liquid thickness above the gold patch center, for three different initial conditions ($0.1$ \text{$\mu$}m, $0.15$ \text{$\mu$}m, $0.2$ \text{$\mu$}m) and compares the evolution of the $0.2$ \text{$\mu$}m curve with $0.15$ \text{$\mu$}m curve and the evolution of $0.15$ \text{$\mu$}m curve with $0.1$ \text{$\mu$}m curve.
The comparison is achieved by translating the $0.1$ \text{$\mu$}m and $0.15$ \text{$\mu$}m curves to the initial values of $0.15$ \text{$\mu$}m and $0.2$ \text{$\mu$}m, respectively; the shifted curves are presented as dashed lines.
The comparison indicates very small difference in liquid evolution.
 
\subsection{Reduced 1D model for $h_{c}(t)$ and phase change as a function of liquid thickness}

\begin{figure}[ht]
\centering
\includegraphics[scale=0.32]{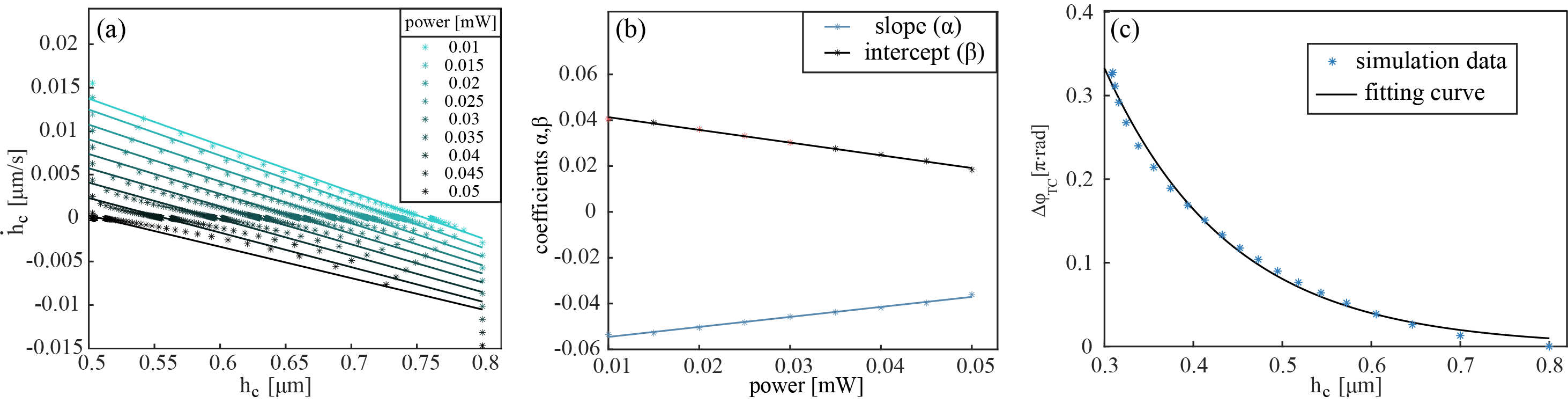}
         \caption{(a) Linear fit evolution in the phase space $(\dot{h},h)$ for different power values together with a linear fit given by Eq.S\ref{ab}. (b) Fitting of the coefficients $\alpha, \beta$ as a function of power.
         (c) The one-to-one relation between liquid thickness $h_{c}$ above gold patch center and the self-induced phase change $\Delta \varphi_{TC}$. 
         The fit (black curve) between the two quantities can be interpolated by $\Delta \varphi_{TC} = 2.786 \cdot \exp \Large( -7.082 \cdot h \Large)$.}
\label{LinearFit}
\end{figure}

In order to incorporate the numerous data curves presented in Fig.S\ref{curves} into a compact numerical model which allows prediction of $h_{c}(t)$ under various driving optical powers, we fit the data to the following first order ordinary differential equation
\begin{equation}
    \dot{h}_{c}(t) = \alpha(P) h_{c}(t) + \beta(P).
\label{ab}    
\end{equation}
Here, $h_{c}(t)$ is liquid thickness above gold patch center as a function of time $t$, $P$ is a constant value of optical power carried by the WG, 
and the fit to simulation data is given in Fig.S\ref{LinearFit}a, and $\alpha(P)$ and $\beta(P)$ are power dependent coefficients explicitly given by
\begin{equation}
    \alpha(P) = 0.4368 \cdot P - 0.0589 ; \quad \beta(P) = -0.5546 \cdot P + 0.0469,
\label{AlphaBeta}    
\end{equation}
and are presented in Fig.S\ref{LinearFit}b.
It is worth mentioning that based on a fitting presented in Fig.S\ref{LinearFit}a, a cubic fit would lead to additional terms in Eq.S\ref{ab} but to small improvement in the model.
We then employ Eq.S\ref{AlphaBeta} to predict liquid evolution $h_{c}(t)$ under power levels $P$ and $2P$ (corresponding to logical 'zero' and 'one') during time window $\tau_{w}$, and zero power during $\tau_{r}$. 
Note that here $P$ stands for the value of the optical power above the WG and not the power in the original input WG; the first coupler divides the power according to $2:1$ ratio ($\tan^2(0.6155) = 1/2$) to compensate for the approximate $3$ db loss due to Joule heating in the gold patch in the liquid cell. 
Consequently, the ratio of input power to the WG power is $3:2$.
For each value of $h_{c}$ we then employ the one-to-one relation between liquid thickness to corresponding phase change which is described in Fig.S\ref{LinearFit}c.

\subsection{Reservoir dynamics: fading memory and delayed 2-bit XOR}

Consider evolution equation Eq.S\ref{ab} with power dependent coefficients $\alpha(P), \beta(P)$ and assume that the power $P$ maintains constant power levels on some time intervals (i.e. piecewise constant), similar to the behavior along $\tau_{W,r}$ intervals considered for XOR task.
Partition the total time period $T$ into $N$ arbitrary intervals at times $[t_{0},t_{1},...,t_{N}]$ of corresponding lengths $[T_{1},...,T_{N}]$ where $t_{N}=T$, and furthermore assume that the liquid thickness at corresponding times is given by $[h_{0},h_{1},...,h_{N}]$.
For any interval where the optical power maintains constant value, the solution to Eq.S\ref{ab} is given by
\begin{equation}
    h(t) = - \dfrac{\beta}{\alpha} + \left( h_{0} + \dfrac{\beta}{\alpha} \right) e^{\alpha t},
\label{absol}    
\end{equation}
where $h_{0}$ is the initial thickness time ($t=0$) and $\alpha, \beta$ are the corresponding values of the coefficients which are constant along the relevant time interval. 
Employing the solution given by Eq.S\ref{absol} for each one of the intervals introduced by the partitioning above, yields
\begin{equation}
    \begin{split}
        h_{1} &= a_{0} + b_{0} h_{0}; \quad a_{0}=\dfrac{\beta_{0}}{\alpha_{0}} \left(e^{\alpha_{0}T_{0}} - 1 \right); b_{0} = e^{\alpha_{0}T_{0}} 
        \\
        h_{2} &= a_{1} + b_{1} h_{1} = a_{1} + b_{1} a_{0} + b_{1} b_{0} h_{0}
        \\
                ...
        \\
        h_{n} &= a_{n-1} + b_{n-1}a_{n-2} + b_{n-1}b_{n-2}a_{n-3}+ ... + b_{n-1}...b_{1}a_{0}+ \underbrace{b_{n-1}...b_{0}h_{0}},
    \end{split}
\label{Iteration}    
\end{equation}
where only the last term, highlighted by the curly bracket, depends on the initial thickness value $h_{0}$.
Here, the constants $a_{k}$, $b_{k}$ are defined by
\begin{equation}
    a_{k}=\dfrac{\beta_{k}}{\alpha_{k}} \left(e^{\alpha_{k}T_{k}} - 1 \right); \quad b_{k} = e^{\alpha_{k}T_{k}}; \quad k = (0,...,n-1),
\end{equation}
where $\alpha_{k}$, $\beta_{k}$ are the values of $\alpha_{P}$ and $\beta_{P}$, respectively, along the time interval $T_{k}$.
Importantly, the last term $h_{0}$ is multiplied by a series of $n$ numbers; each number $b_{k}$ ($k=(0,...,n-1)$) is smaller than unity because $\alpha(P)<0$.
Consequently, for sufficiently large number of steps $n$ the last term will become arbitrary close to zero and therefore negligible compared to $a_{n-1}$ and other terms.
\begin{figure}[h!]
\centering
\includegraphics[scale=0.38]{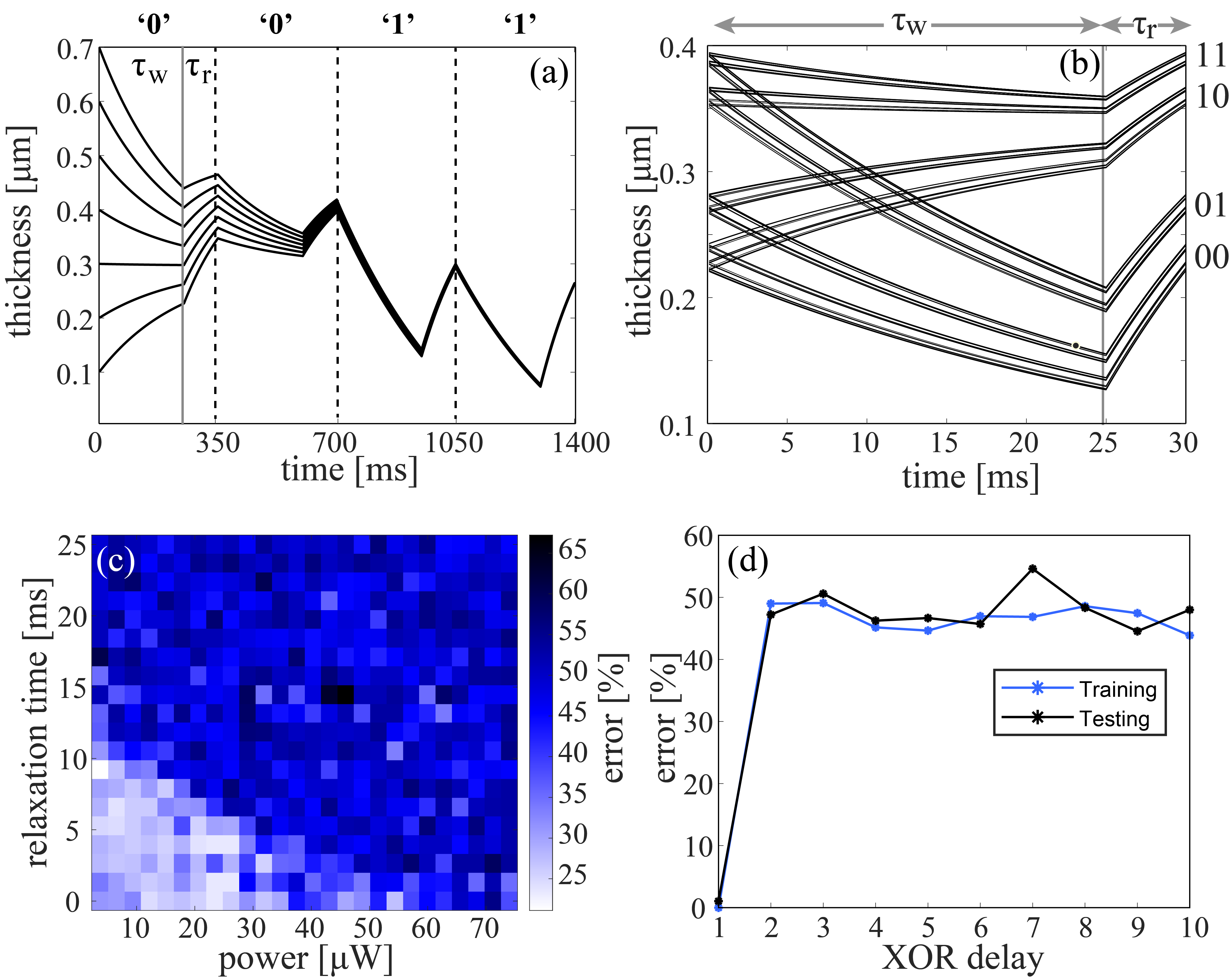}
         \caption{Numerical results presenting: (a) illustration of the fading memory effect, where the evolving system 'forgets' the different initial conditions leading to convergence towards a common attractor state; (b) folded thickness dynamics diagram where thickness evolution in nonlocal circuit is divided to time intervals of length $T= \tau_{w} + \tau_{r}$ and plotted on top for each other between times $5T$ to $200T$. 
         The diagram shows four main curves demonstrating four different combinations due to two possible actuation pulses and one step back in time dominant memory (key parameters: $\tau_{w}=25$ ms, $\tau_{r}=5$ ms, $P=0.033$ mW); (c) test results of XOR gate with delay $d=2$ ($\tau_{w} = 25$ ms, $\tau_{r} = 10$ ms). (d) Error performance of XOR operation as a function of delay $d$ for the following parameters: $\tau_{r}=10$ ms, $\tau_{w}= 25$ ms, $P=0.06$ mW. }
\label{XORd}
\end{figure}
Fig.S\ref{XORd}a presents the fading memory effect where the evolving thin liquid film, governed by Eq.S\ref{ab}, indeed 'forgets' the initial conditions as we derived in Eq.S\ref{Iteration}, and all curves approach towards a common attractor state.
Fig.S\ref{XORd}b presents test results of 2-bit XOR with delay $d=2$ as a function of power and rest time ($\tau_{r}$); as expected it admits lower accuracy compared to the $d=1$ case presented in Fig.5.
Furthermore, Fig.S\ref{XORd}c presents test error of 2-bit XOR as a function of delay ($d$) for fixed power parameters thus illustrating that $d=1$ can achieve minimal error and that generally other values of $d$ lead to low accuracy.

\subsection{Nonlocal RC: relation between liquid thickness to the output light intensity}

\begin{figure}[h!]
\centering
\includegraphics[scale=0.45]{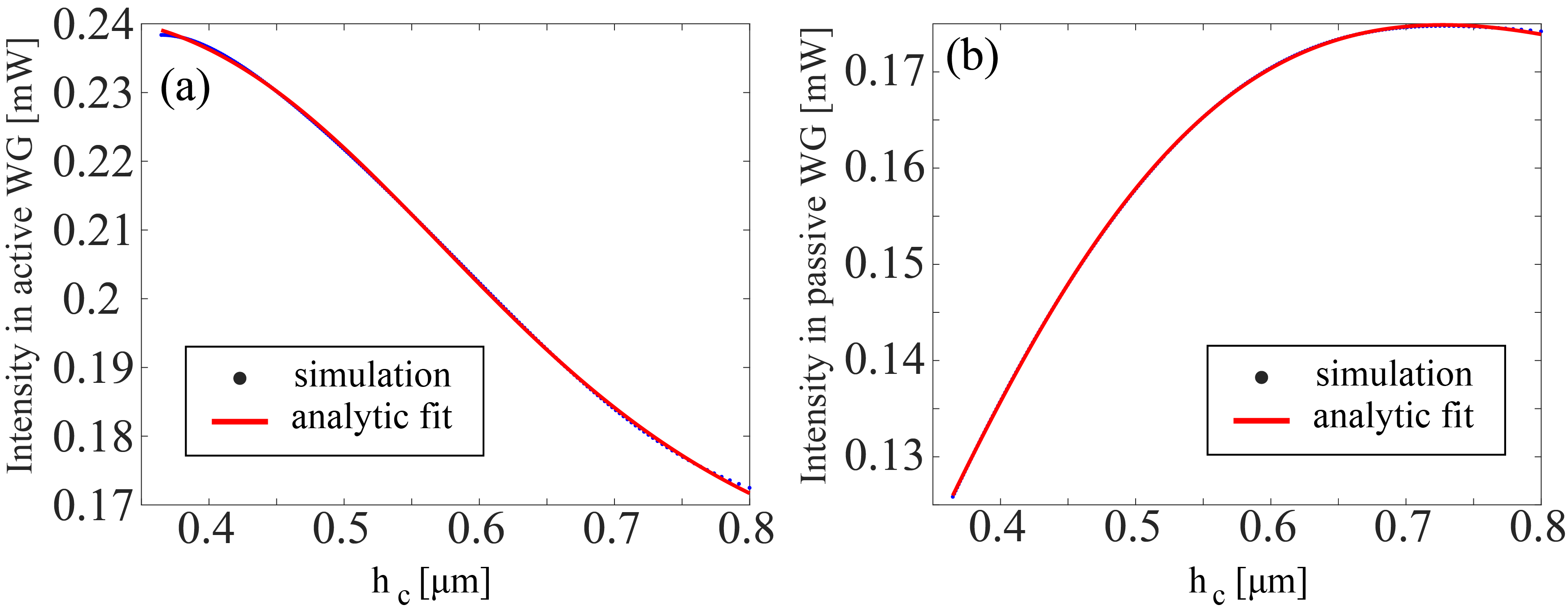}
         \caption{Multiphysics simulation result describing dependence  
         of optical output power in active (a) as well as (b) passive WGs
         in a directional coupler geometry  as a function of liquid film thickness, where the two parallel WGs have separation distance $500$ nm (see Fig.\ref{XOR}a).
         The analytic fit, given by Eq.S\ref{CouplingVsIntensityEq}, allows to employ this result in RC simulation.}
\label{CouplingVsIntensity}
\end{figure}

In this section we consider linear coupler geometry where the coupling region is covered with liquid. The heating in the gold patch in active WG leads to thickness change, which in turn modifies the coupling coefficient and the intensity in the output ports. Fig.S\ref{CouplingVsIntensity}a presents the corresponding geometry whereas Fig.S\ref{CouplingVsIntensity}b,c present intensity in the active and passive WGs, respectively.
The simulation points is fitted against a simple analytical curve, given by
\begin{equation}
\begin{split}
 \text{active WG} = (0.175 \cdot h_{c}^{2} - 0.197 \cdot h_{c} + 0.088)/(h_{c}^{2} - 0.978 \cdot h_{c} + 0.386)
    \\
 \text{passive WG} = (0.093 \cdot h_{c}^{2} + 0.005 \cdot h_{c} + 0.014)/(h_{c}^{2} - 0.66 \cdot h_{c} + 0.326),
\end{split}
\label{CouplingVsIntensityEq}
\end{equation}
Eq.S\ref{CouplingVsIntensityEq} 
thus allowing simpler implementation of nonlocal RC.

\subsection{Comparison against optical-based and liquid-based RC platforms}

In the following Table S\ref{Comparison} we compare our Optofluidic concept with thin liquid film, against several other integrated chip-scale photonics approaches which include Phase change approach which relies on phase change transition, All-photonic approach which relies on optical delay without material response, as well as other modalities which rely on external modulation/feedback due to Electro-optic and Thermo-optic effects.
For completeness we also provide Fluid-mechanical approach without optical interaction, which employs mechanical actuation of surface waves in liquid and their reflection from the large container boundaries as an internal feedback which can be employed for RC/NC computation.

Among the different criteria used to conduct the comparison we choose the magnitude of the corresponding refractive index change; 
if the feedback was realized by external unit, delay or was it induced by internal nonlinear material response;
size of the active area where the material response triggers induced nonlinear changes of the refractive index; energy per bit.
\begin{table}[h]
\centering
\begin{tabular}{ |p{3.0cm}||p{1.4cm}||p{2.1cm}||p{1.6 cm}||p{1.7cm}||p{1.4cm}||p{1.3cm}||p{1.2cm}||p{0.7cm}|}
 \hline
   & Index change  & Area & Feedback & Time scale & Non- linearity  &  Non- locality & Energy per bit & Ref \\
 \hline
 Electro-optic   & O$(10^{-4})$ & $10 \times 10$ \text{$\mu$}m$^{2}$ & external & O$(100$ ps$)$ & \hspace{0.22in}$\times$ & \hspace{0.2in}$\times$ & $500$ fJ & [45] \\
  \hline
 Thermo-optic & O$(10^{-3})$ & $4 \times 5$ \text{$\mu$}m$^{2}$ & 
 external & 
 O$(100$ \text{$\mu$}s$)$ & 
 \hspace{0.19in} $\times$ &
 \hspace{0.17in} $\times$ & 
 $0.4$ nJ & 
 here \\
  \hline
     All-photonic & \hspace{0.22in} -	& $16$ mm$^{2}$ 	& 
 delay & 
 O$(300$ ps$)$ & 
 \hspace{0.20in} $\times$ & 
 \hspace{0.18in} $\times$ & 
 $0.1$ fJ & 
 [14] \\
  \hline
 Phase change & O$(10^{-2})$ & $0.4 \times 0.4$ \text{$\mu$}m$^{2}$ & 
 \textbf{internal}	& 
 O$(100$ ns$)$	& 
 \hspace{0.22in}\checkmark	& 
 \hspace{0.22in}$\times$ & $400$ pJ & 
 [47] \\
  \hline
 Fluid-mechanical & \hspace{0.22in} -	& $1 \times 1$ m$^{2}$ 	& 
 \textbf{internal} & 
 O$(10$ s$)$ & 
 \hspace{0.22in} - & 
 \hspace{0.22in} - & $0.3$ J & 
 [43] \\
  \hline
 Optofluidic & O$(10^{-1})$  & 
 $0.5 \times 0.5$ \text{$\mu$}m$^{2}$ & 
 \textbf{internal} & 
 O$(10$ ms$)$ & \hspace{0.2in} \checkmark & \hspace{0.2in} \checkmark & $10$ nJ & here \\
  \hline
 \hline
\end{tabular}
\caption{Comparison between several approaches capable in-principle to realize NC/RC tasks using integrated photonics chip-scale platforms, with internal and external feedback as main criteria which strongly affects the footprint of the photonic device. 
Electro-optic and Thermo-optic based approaches allow relatively efficient signal modulation of on-chip component in terms of energy (Electro-optic also in time), but require external control unit and hence additional space/energy resources.
Phase change and Optofluidic approaches admit internal feedback due to matter response in a compact region, which results in a smaller footprint without optical feedback lines, yet these are not as fast as Electro-optic.
All-photonic approach admits built-in photonic delay lines and requires very small operation energy, yet photonic delay lines translate into larger footprint compared to other chip-scale approaches. 
The energy per-bit in all cases is computed without considering the power needed to operate the corresponding detection system, and in Electro-optic and Thermo-optic approaches it also does not take into account the circulating optical energy in the circuit, which is lower compared to the corresponding values required to modulate the device.
The macroscopic Fluid-mechanical approach is brought here for reference; it admits internal feedback due to reflection of water waves from container's boundaries, yet the macroscopic size of the mechanical actuators leads to meter sized system with much higher energy consumption.}
\label{Comparison}
\end{table}
Here, the Optofluidic approach presents the highest index change due to the large refractive index contrast between gas and liquid media, which presumably leads to high signal modulation and to higher performance of the corresponding NC/RC tasks.
Importantly, while the Electro-optic and Thermo-optic approaches, require memorizing previous measurement results in external memory unit and introducing external feedback into the photonic circuit through electrical modulation of micro-ring resonator and metal heater in MZI device (shortly discussed below), respectively, Optofluidic and Phase-change approaches rely on internal feedback due to nonlinear material response and transient evolution of the excited state towards equilibrium which is employed as a memory at the same spatial location.
For instance, in the Optofluidic approach the optically triggered liquid response allows to modulate the signal and also to serve as an optical memory without the need to introduce dedicated electrical/optical feedback loops.
The corresponding energy per bit for the digital XOR task is given by $10$ \text{$\mu$}W $\times$ $1$ ms $=10$ nJ/bit (where $10$ text{$\mu$}W is the energy allowing to support high performance as read from diagram Fig.\ref{XOR}e), which is used for both optical transmission and actuation of nonlinear liquid response.
Analogously, assuming MZI geometry similar to Fig.\ref{XOR}a but with a metal heater approximately two microns above the WG, can lead to an optical phase shift if the temperature in the heater is externally modulated.
In particular, assuming WG temperature increase is $20^{o}$ which requires metal patch increase $\Delta T = 40^{o}$ degrees in order to achieve $\pi/2$ phase shift (see for instance for Applied Nanotools foundry test results https://www.appliednt.com/nanosoi/sys/resources/examples/tri-layer/), implies that the total heat energy is given by
$q = \rho \cdot V \cdot c \cdot \Delta T \approx 7.5$ nJ. 
Here, we assumed parameters corresponding to integrated titanium-tungsten (TiW) heaters with density $\rho = 4429$ kg/m$^{3}$, specific heat $c=526.3$ J/(kg $\cdot$ K), and volume $V=0.2 \cdot 4 \cdot 100 = 80$ \text{$\mu$}m$^{3}$ of the $100$ \text{$\mu$}m long metal patch.
To achieve smaller modulation by a factor of $20$, comparable to modulation of Optofluidic circuit presented in Fig.S\ref{CouplingVsIntensity} which is sufficient for XOR implementation, requires shorter metal patch of length $100$\text{$\mu$}m$/20=5$ \text{$\mu$}m and correspondingly smaller modulation energy $7.5/20 \approx 0.3$ nJ. 
The All-photonic approach on the other hand, presents an example of chip-scale system which enables to obtain a linear response (in electric field) due to delay lines which provide the capability to form interference between signals injected at different times, but it does not have built-in nonlinear matter response apart the square law detection.
While in Phase change and Optofluidic modalities the nonlinearity stems from light-matter interaction between WG modes to phase change substance or liquid, in the Fluid-mechanical approach the nonlinearity stems from inherently nonlinear evolution of the gas-liquid interface; 
linear evolution occurs in a small amplitude regime.
Interestingly, while optical nonlinearity in the phase change approach is limited to a single WG, Optofluidic approach is the only one which allows to harness nonlocality for computation and even demonstrates superior performance compared to nonlinear effect (see Fig.\ref{XOR},\ref{01}). 


\subsection{Analog task: row by row data injection}

In Fig.\ref{01} we presented performance of nonlinear and nonlocal circuits, schematically presented in Fig.\ref{XOR}(a,c), respectively, to perform the analog task of handwritten 'zero'/'one' classification based on 'bit-after-bit' parallel injection in two input arms, where the signal of each bit is proportional to the brightness value of the corresponding pixel in the image.
\begin{figure}[h!]
\centering
\includegraphics[scale=0.55]{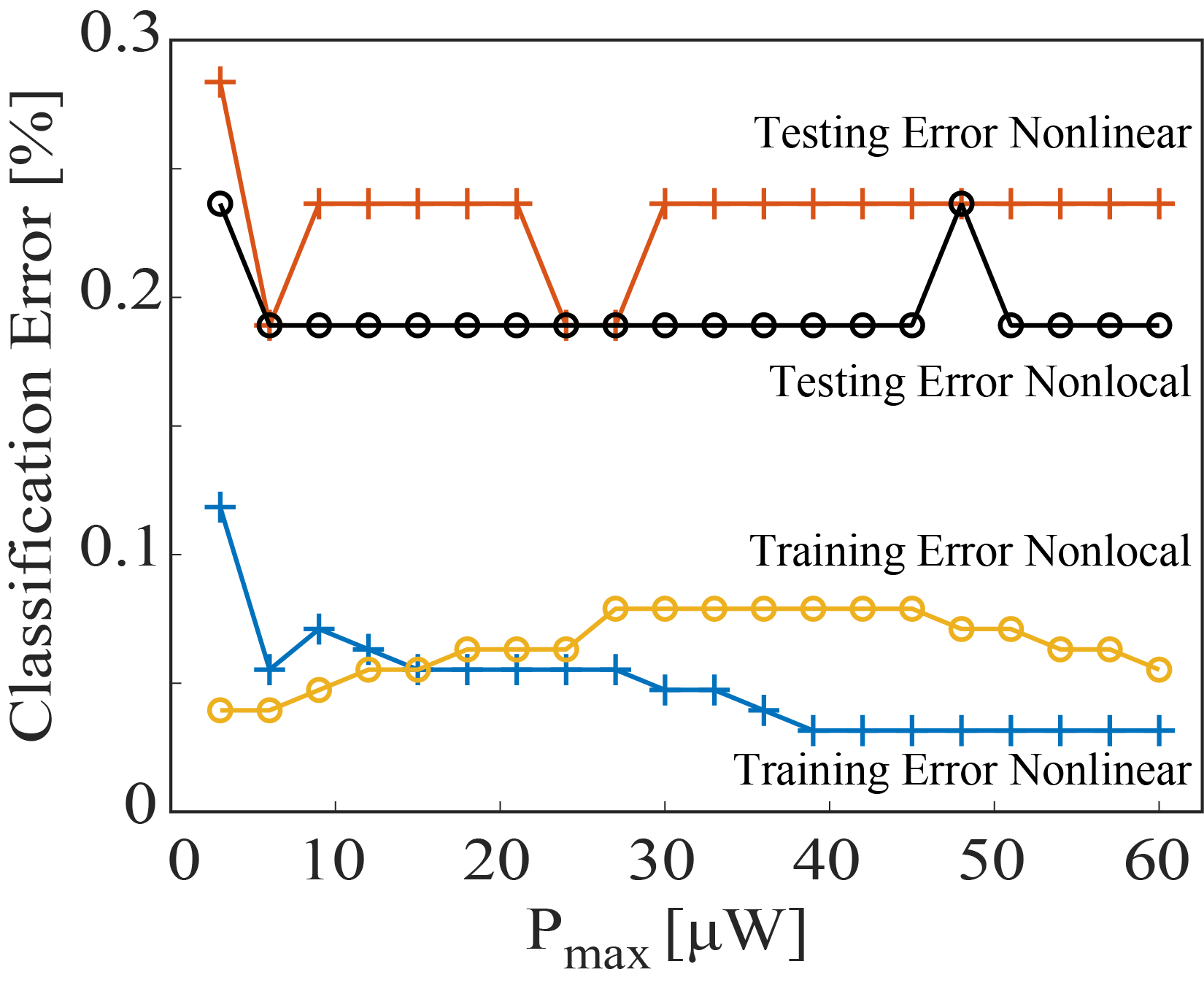}
         \caption{Handwritten 'zero'/'one' classification based on parallel columns injection into nonlinear and nonlocal circuits. The two sets of curves marked with plus and circle correspond to performance of nonlinear and nonlocal circuits, respectively.}
\label{01_columns}
\end{figure}
Here, we present how the classification performance changes if one simultaneously injects a pair of columns instead of rows.
Fig.S\ref{01_columns} presents simulation results where the nonlinear and nonlocal circuits perform classification of handwritten 'zero'/'one' digits by employing parallel 'bit-after-bit' injection of two pairs of columns ($28$ bits long each); one column is injected in each one of the two input arms, similarly to the row injection scheme.
We see that overall the performance is comparable to rows injection presented in Fig.\ref{01}, indicating that the degree of correlation generated between adjacent rows/columns is similar in the type of basic circuits we considered.
While optimizing the injection scheme and constructing optimized task-dependent circuits is an important question, it is beyond the scope of this work.

\subsection{Optofluidic RNN}

While the circuits presented in Fig.\ref{XOR} employ single or two input WGs, it should require substantial computation time for task which admit large number of bits, including 'zero'/'one' handwritten classification task presented above.
To achieve time-division multiplexing one can naturally employ more input waveguides each carrying independent information.
\begin{figure}[h!]
\centering
\includegraphics[scale=0.45]{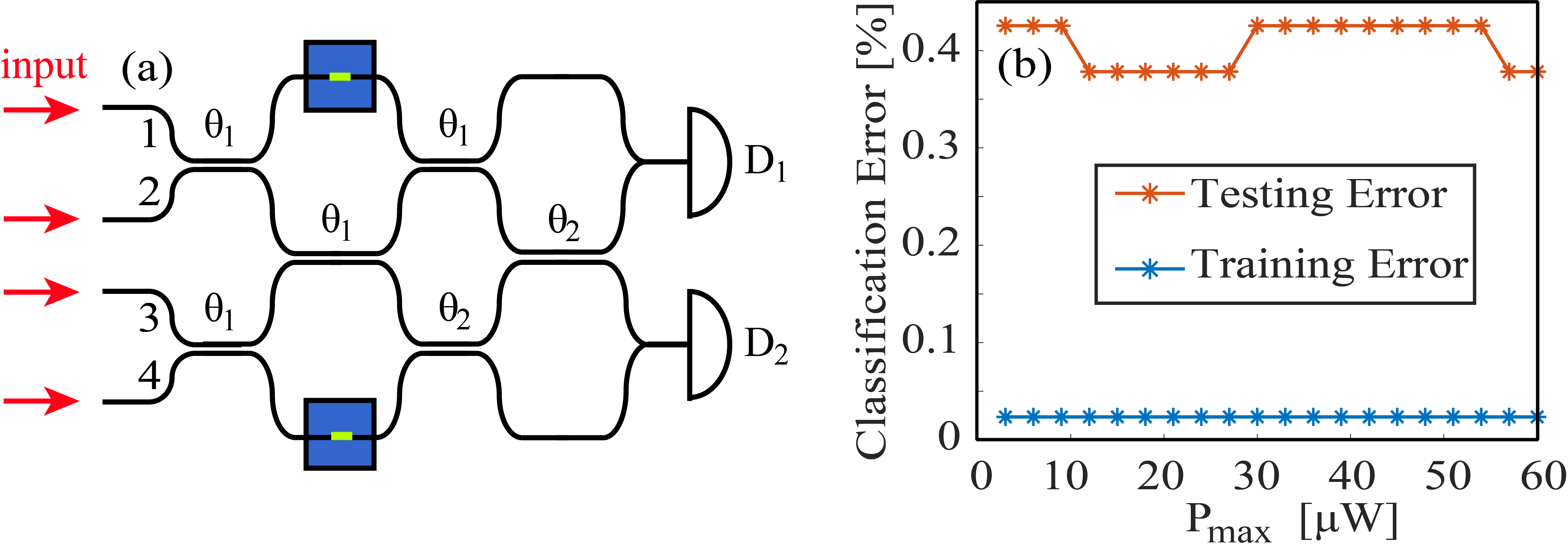}
         \caption{Optofluidic RNN with internal feedback processing handwritten 'zero'/'one' classification.
         (a) Schematic representation of the network with four inputs and two liquid cells, and (b) the corresponding training and test results presenting the corresponding classification error.}
\label{Network}
\end{figure}
Here, we present a basic example towards this direction by employing four inputs which decrease the computation time by a factor of two compared to the schemes presented in Fig.\ref{XOR}(a,c).    
Fig.S\ref{Network} presents an optofluidic network comprised of four inputs, pair liquid cells, six linear couplers (with coupling coefficients $\theta_{1,2}$), and a pair of optical detectors $D_{1,2}$. 

Each input WG in scheme Fig.S\ref{Network}a, labeled as $1,2,3,4$ receives a different line in the image according to $(4i-3,4i-2,4i-1,4i)$ where the $1^{st}$/$2^{nd}$/$3^{rd}$/$4^{th}$ entry in this vector corresponds to the input fed into WG $1,2,3,4$, respectively, and $i=(1,..,N/4)$ where $N=28$ is the total number of rows in the image. 
Fig.S\ref{Network}b presents training and test error of 'zero'/'one' classification with similar performance compared to single MZI optofluidic circuit results presented in Fig.\ref{01}, but allowing twice faster computation. 
Also, since the area of significantly deformed gas-liquid interface around the gold patch can be limited to only few microns or less (e.g. if the actuation does not operate for a long time), in principle larger number of WGs can be accommodated within a single liquid cell if cross talk nonlocal effects are not desired.

\subsection{Preparatory experiments: liquid deposition and plausibility of the approach}

\begin{figure}[h!]
\centering
\includegraphics[scale=0.37]{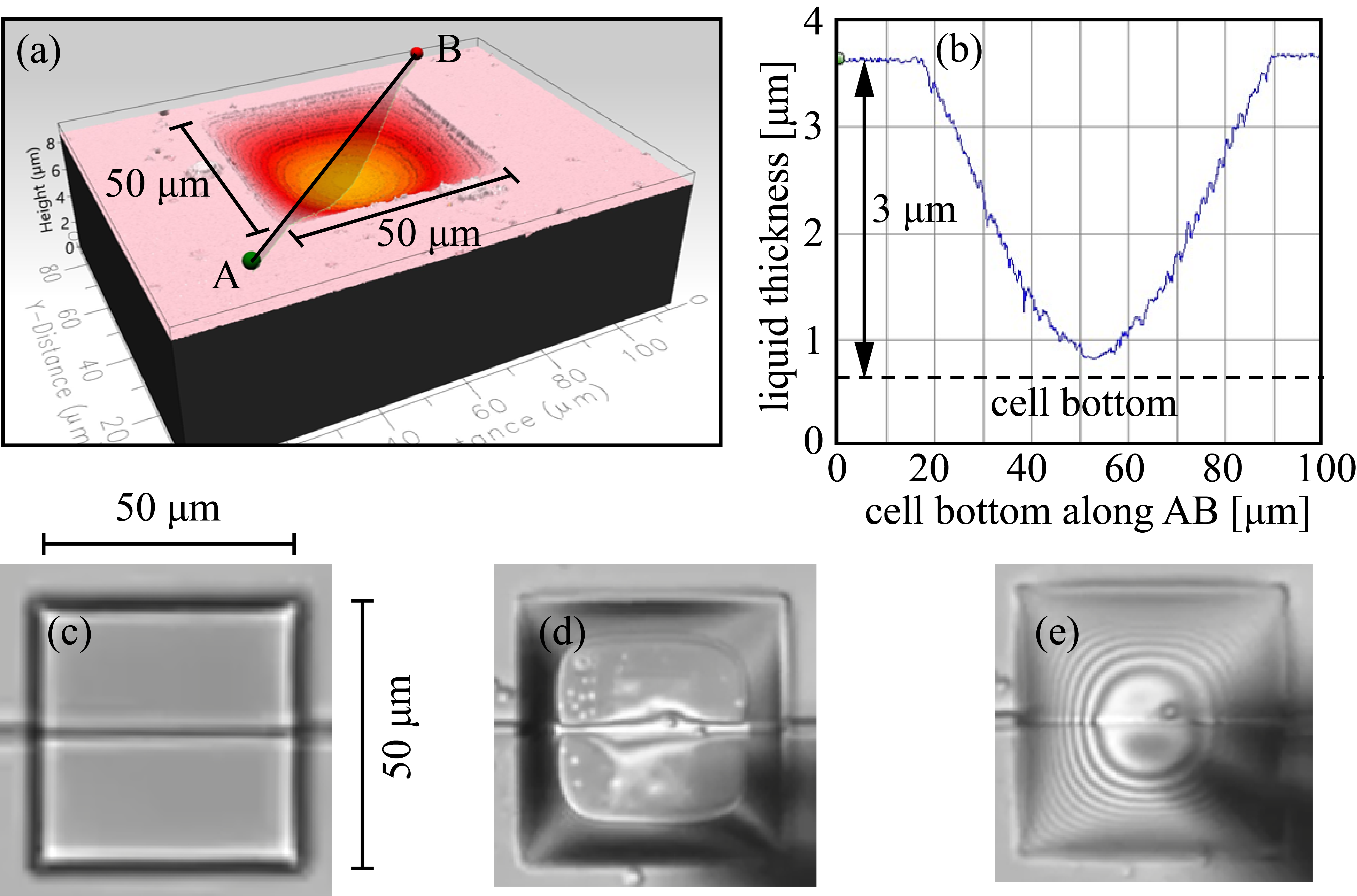}
         \caption{Experimental images of silicone oil deposited into an etched $50$ $\times$ $50$ $\times$ $3$ \text{$\mu$}m$^{3}$ cell in buried oxide (BOX) layer. (a) 3D images presenting smooth oil surface, (b) oil film thickness along the diagonal AB, (c-e) top view presenting empty, partially filled, and full cell, respectively.
         The central horizontal line is a $500$ nm wide silicon WG, which does not interfere with the process of controlled liquid deposition.}
\label{Deposition}
\end{figure}

To demonstrate feasibility of our approach from experimental point of view, especially controlling liquid deposition on a micron scale, we fabricated trenches in silicone chip and then deposited silicone oil into it.
Reactive-ion etching (RIE) was used to etch the silicon WG, RIE and Buffered oxide etch (BOE) were used to remove the cladding silicon dioxide till buffer silicon dioxide layer without damaging the WG.
Fig.S\ref{Deposition}a, which was generated by using Profilm 3D$^\circledR$ (Filmetrics, San Diego, CA, USA), presents 3D silicone oil profile within square-shaped $50$ \text{$\mu$}m cell wide and $3$ \text{$\mu$}m deep,
where the refractive index of silicone oil has the same refractive index as silicon dioxide in visible wavelength, so the 3D profile of the tranche filled with liquid could be generated by white light interferometry (WLI) method.
Fig.S\ref{Deposition}b presents the profile of the gas-liquid interface along the diagonal AB.
Fig.S\ref{Deposition}(c-e) presents top view microscopy image of the $50$ \text{$\mu$}m cell hosting a WG (central horizontal line) and deposited liquid at three different stages: (c) empty, (d) partially filled, (e) fully filled.
Note that due to wetting of the vertical walls, liquid thickness is thinner at the center reaching to sub micron thickness which according to our model is enough to initiate self-induced film deformation.

Note that in our modeling (e.g. Fig.2) we assumed initially flat gas-liquid interface, whereas in practice Fig.S\ref{Deposition} presents curved interface with increasingly higher thickness as liquid film approaches $3$ \text{$\mu$}m high vertical walls, and thin liquid film of minimal thickness $\sim 200$ nm in the central region above the WG. 
Nevertheless, the average slope of the interface is only $\sim (3-0.2)⁄25 \approx 0.11$ (where $25$ \text{$\mu$}m denotes the minimal distance from cell's center to its boundary), and the WG is located at region where the liquid film has a minima point and hence the local slope is even smaller. 

Our previous experiments where thin liquid film was deposited on gold plasmonic grating \cite{RubinHong2019}, indicates that thin liquid thinning can be invoked by external light illumination across a wide range of power levels. For instance $\sim 400$ mW can initiate large deformation which lead to liquid depletion (holes), whereas $\sim 6$ \text{$\mu$}W leads to subnanometer thickness changes.
Importantly, in all cases the liquid film restored to initial configuration under surface tension forces.
Since our modeling predicts that power levels of just around $0.1$ mW (see Fig.2) are enough in order to achieve sufficiently high temperature increase and liquid deformation due to higher power density in the channel WGs, we believe that the proposed approach is promising in terms of future experimental realization.  

\subsection{Self-induced absorption in the metal patch and the waveguide}

\begin{figure}[h!]
\centering
\includegraphics[scale=0.48]{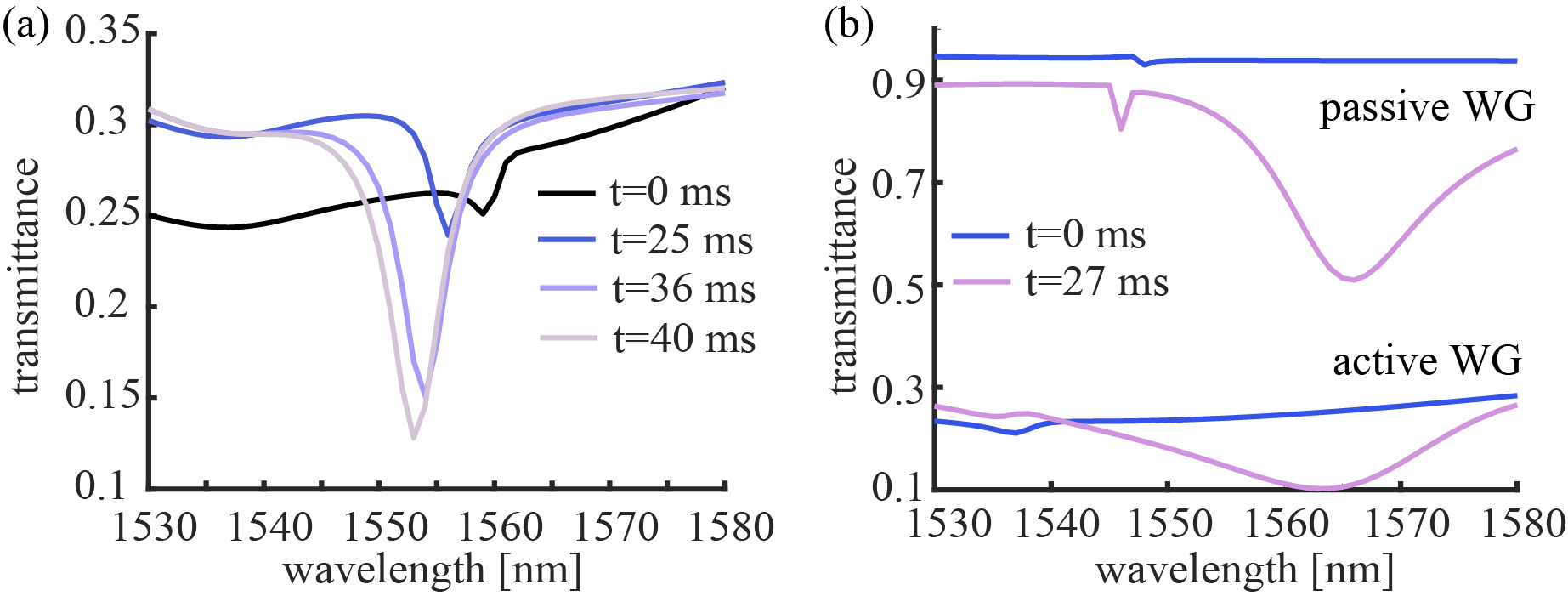}
         \caption{Multiphysics simulation results presenting transmittance as a function of wavelength for different moments of time under CW $0.07$ mW TM mode. (a) Single active WG described by Fig.2, (b) passive WG described by Fig.3. Note that for the passive WG at our working wavelength $1550$ nm, only $\sim 0.25$ dB loss is induced by the gold patch and deformation of the thin liquid film.}
\label{resonance}
\end{figure}

\textcolor{black}{Here, we investigate the effect of the gold patch on transmittance of the photonic modes as a function of mode's wavelength. 
Fig.S\ref{resonance} presents multiphysics simulation results of transmittance as a function of wavelength and of time in the spectral region between $1530$ nm to $1580$ nm, due to CW TM mode of power $0.07$ mW, where Fig.S\ref{resonance}a corresponds to the case of active WG described in Fig.2 whereas Fig.S\ref{resonance}b corresponds to the case of active and passive WGs described in Fig.3.
In particular, Fig.S\ref{resonance}a describes the case of time dependent transmittance spectra where $t=0$ ms corresponds to initially flat liquid film, whereas progressively later time moments with more prominent liquid deformation reveal emergence of transmittance dip around $1553$ nm at $t=40$ ms indicating higher quality factor and resonant absorption.
}

\end{document}